\documentclass[fleqn,usenatbib]{mnras}

\usepackage[T1]{fontenc}

\DeclareRobustCommand{\VAN}[3]{#2}
\let\VANthebibliography\thebibliography
\def\thebibliography{\DeclareRobustCommand{\VAN}[3]{##3}\VANthebibliography}

\usepackage{graphicx}	
\usepackage{amsmath}	
\usepackage{comment}
\usepackage{txfonts}
\usepackage{hyperref}
\usepackage{textgreek}
\usepackage{xargs}
\usepackage{xspace}
\usepackage{booktabs} 
\usepackage[dvipsnames,hyperref]{xcolor}

\usepackage{hyperref}
\hypersetup{
    colorlinks=true,
    linkcolor=Cerulean,
    citecolor=NavyBlue,
    filecolor=magenta,
    urlcolor=Violet,
    pdftitle= {JADES_gsz11.2},
    pdfpagemode=FullScreen,
    }

\newcommand{\Msun}{\ensuremath{\mathrm{M}_\odot}\xspace}
\newcommand{\jwst}{\textit{JWST}\xspace}

\newcommandx{\permittedEL}[6][1=O,2=III,3=,4=,5=,6=]{\text{{#1}\,{\sc {#2}}{#3}{#4}{#5}{#6}}\xspace}
\newcommandx{\semiforbiddenEL}[6][1=O,2=III,3=,4=,5=,6=]{\text{{#1}\,{\sc {#2}}]{#3}{#4}{#5}{#6}}\xspace}
\newcommandx{\forbiddenEL}[6][1=O,2=III,3=,4=,5=,6=]{\text{[{#1}\,{\sc{#2}}]{#3}{#4}{#5}{#6}}\xspace}
\newcommand{\kms}{km s$^{-1}$}

\newcommand{\HeIIL}[1][1=1640]{\permittedEL[He][ii][\textlambda][#1]}

\newcommandx{\OIIL}[1][1=3728]{\forbiddenEL[O][ii][\textlambda][#1]}
\newcommand{\OIIall}{\forbiddenEL[O][ii][\textlambda][\textlambda][3726,][29]}
\newcommand{\OIII}{\forbiddenEL[O][iii]}
\newcommandx{\OIIIL}[1][1=5007]{\forbiddenEL[O][iii][\textlambda][#1]}
\newcommand{\OIIIall}{\forbiddenEL[O][iii][\textlambda][\textlambda][5007,][4959]}

\newcommandx{\NIIL}[1][1=6585]{\forbiddenEL[N][ii][\textlambda][#1]}

\newcommand{\NeIII}{\forbiddenEL[Ne][iii]}
\newcommandx{\NeIIIL}[1][1=3869]{\forbiddenEL[Ne][iii][\textlambda][#1]}

\newcommand{\CIIIall}{\semiforbiddenEL[C][iii][\textlambda][\textlambda][1907,][09]}
\newcommand{\CIVall}{\permittedEL[C][iv][\textlambda][\textlambda][1548,][51]}
\newcommand{\OIIIUvall}{\semiforbiddenEL[O][iii][\textlambda][\textlambda][1661,][66]}
\newcommand{\CII}{\forbiddenEL[C][ii][\textlambda][158$\mu$m]}

\newcommand{\NIVall}{\semiforbiddenEL[N][\textsc{iv}][\textlambda\textlambda][1483,][86]}
\newcommand{\NIIIall}{\semiforbiddenEL[N][\textsc{iii}][\textlambda][1748-][54]}

\newcommand{\NIV}{\semiforbiddenEL[N][\textsc{iv}]}
\newcommand{\NIII}{\semiforbiddenEL[N][\textsc{iii}]}

\newcommand{\NeIV}{\forbiddenEL[Ne][iv]}
\newcommand{\NeV}{\forbiddenEL[Ne][v]}

\newcommand{\target}{GS-z11-1\xspace}

\newcommand{\JWST}{\textit{JWST}\xspace}

\newcommand{\ticks}[1]{%
      \reflectbox{'}#1'}
\newcommand{\beagleagn}{\textsc{beagle-agn}\xspace}
\newcommand{\beagle}{\textsc{beagle}\xspace}
\newcommand{\multinest}{\textsc{MultiNest}}
\newcommand{\txn}[1]{\textnormal{#1}}
\newcommand{\HII}{\mbox{H\,{\sc ii}}}
\newcommand{\Z}{\hbox{$\txn{Z}$}}
\newcommand{\yr}{\hbox{$\txn{yr}$}}
\newcommand{\Lacc}{\hbox{$L_\txn{acc}$}}
\newcommand{\PLalpha}{\hbox{$\alpha_\textsc{pl}$}}

\newcommand{\logUsAGN}{\hbox{$\log U_\textsc{s}^\textsc{nlr}$}}

\newcommand{\xidAGN}{\hbox{$\xi_\txn{d}^\textsc{nlr}$}}
\newcommand{\ZAGN}{\hbox{$\Z_\txn{gas}^\textsc{nlr}$}}

\newcommand{\tauV}{\hbox{$\hat{\tau}_\textsc{v}$}}

\newcommand{\xid}{\hbox{$\xi_\txn{d}$}}
\newcommand{\Zism}{\hbox{$\Z_\textsc{ism}$}}
\newcommand{\Zhii}{\hbox{$\Z_\txn{gas}^\textsc{hii}$}}
\newcommand{\Zsun}{\hbox{$\Z_{\odot}$}}

\newcommand{\sfrconsistent}{\hbox{log(SFR/M$_{\odot}$/yr)}}

\newcommand{\MtotInLog}{\hbox{$\txn{M}_\txn{tot}$}}
\newcommand{\tausfr}{\hbox{$\tau_\textsc{sfr}$}}
\newcommand{\logUs}{\hbox{$\log U_\textsc{s}$}}
\newcommand{\logUsHII}{\hbox{$\log U^{\textsc{HII}}_\textsc{s}$}}
\newcommand{\Mstar}{\hbox{$M_{\star}$}}

\newcommand{\msun}{\ensuremath{\mathrm{M_\odot}}\xspace}

\title[GS-z11-1]{JADES: Carbon-enhanced, Nitrogen-normal compact galaxy at z=11.2}

\author[Jan Scholtz]{\parbox[h]{\textwidth}{
J. Scholtz$^{1,2}$, M. S. Silcock$^{3}$, E. Curtis-Lake$^{3}$, R. Maiolino$^{1,2,4}$, S. Carniani$^{5}$, F. D'Eugenio$^{1,2}$, X. Ji$^{1,2}$,  P. Jakobsen$^{6,7}$, K. Hainline$^{8}$, S. Arribas$^{9}$, W. M. Baker$^{10}$, R. Bhatawdekar$^{11}$, A. J. Bunker$^{12}$, S. Charlot$^{13}$, J. Chevallard$^{12}$, M. Curti$^{14}$, Daniel J.\ Eisenstein$^{15}$, Y. Isobe$^{1,2,16}$, G. C. Jones$^{1,2}$, E. Parlanti$^{5}$, P. G. Pérez-González$^{9}$, P. Rinaldi$^{8}$, B. Robertson$^{17}$, S. Tacchella$^{1,2}$, H. \"Ubler$^{18}$, C. C. Williams$^{19}$ , C. Willott$^{20}$, J. Witstok$^{6,7}$
}\vspace{0.4cm}
\\
$^{1}$Kavli Institute for Cosmology, University of Cambridge, Madingley Road, Cambridge, 
CB3 0HA, UK\\
$^{2}$Cavendish Laboratory, University of Cambridge, 19 JJ Thomson Avenue, Cambridge CB3 0HE, UK\\
$^{3}$Centre for Astrophysics Research, Department of Physics, Astronomy and Mathematics, University of Hertfordshire, Hatfield AL10 9AB, UK \\
$^{4}$Department of Physics and Astronomy, University College London, Gower Street, London WC1E 6BT, UK\\
$^{5}$ Scuola Normale Superiore, Piazza dei Cavalieri 7, I-56126 Pisa, Italy\\
$^{6}$ Cosmic Dawn Center (DAWN), Copenhagen, Denmark\\
$^{7}$ Niels Bohr Institute, University of Copenhagen, Jagtvej 128, DK-2200, Copenhagen, Denmark\\
$^{8}$ Steward Observatory, University of Arizona, 933 N. Cherry Avenue, Tucson, AZ 85721, USA\\
$^{9}$ Centro de Astrobiolog\'ia (CAB), CSIC–INTA, Cra. de Ajalvir Km.~4, 28850- Torrej\'on de Ardoz, Madrid, Spain\\
$^{10}$ DARK, Niels Bohr Institute, University of Copenhagen, Jagtvej 155A, DK-2200 Copenhagen, Denmark\\
$^{11}$ European Space Agency (ESA), European Space Astronomy Centre (ESAC), Camino Bajo del Castillo s/n, 28692 Villanueva de la Cañada, Madrid, Spain\\
$^{12}$ University of Oxford, Department of Physics, Denys Wilkinson Building, Keble Road, Oxford OX13RH, United Kingdom\\
$^{13}$ Sorbonne Universit\'e, CNRS, UMR 7095, Institut d'Astrophysique de Paris, 98 bis bd Arago, 75014 Paris, France\\
$^{14}$ European Southern Observatory, Karl-Schwarzschild-Strasse 2, 85748 Garching, Germany\\
$^{15}$ Center for Astrophysics $|$ Harvard \& Smithsonian, 60 Garden St., Cambridge MA 02138 USA\\
$^{16}$ Waseda Research Institute for Science and Engineering, Faculty of Science and Engineering, Waseda University, 3-4-1, Okubo, Shinjuku, Tokyo 169-8555, Japan\\
$^{17}$ Department of Astronomy and Astrophysics, University of California, Santa Cruz, 1156 High Street, Santa Cruz, CA 95064, USA\\
$^{18}$ Max-Planck-Institut f\"ur extraterrestrische Physik (MPE), Gie{\ss}enbachstra{\ss}e 1, 85748 Garching, Germany\\
$^{19}$ NSF National Optical-Infrared Astronomy Research Laboratory, 950 North Cherry Avenue, Tucson, AZ 85719, USA\\
$^{20}$ NRC Herzberg, 5071 West Saanich Rd, Victoria, BC V9E 2E7, Canada\\
}

\date{Accepted XXX. Received YYY; in original form ZZZ}

\pubyear{2025}

\begin{document}
\label{firstpage}
\pagerange{\pageref{firstpage}--\pageref{lastpage}}
\maketitle

\begin{abstract}

Over the past few years \textit{JWST} has been a major workhorse in detecting  and constraining the metal enrichment of the first galaxies in the early Universe and finding the source of the ionisation of their interstellar medium. In this work, we present new deep JWST/NIRSpec spectroscopy of GS-z11-1, a galaxy at z = 11.28, in which we report the detection of multiple rest-frame UV and optical emission lines: CIII]$\lambda\lambda$1907,09, CIV]$\lambda\lambda$1548,51, [OII]$\lambda\lambda$3726,29, [NeIII]$\lambda$3869, H$\gamma$ and tentative evidence for HeII$\lambda$1640.
The ionisation properties of GS-z11-1 are consistent with star formation, with potential contribution from an active galactic nucleus (AGN).
We estimate a galaxy stellar mass of log($\Mstar$/M$_{\odot}$) = 7.8$\pm$0.2 and log(SFR/(M$_{\odot}$ yr$^{-1}$))= 0.32$\pm$0.11 for the fiducial SF-only models. We measured C/O from the SED modelling of C/O = 1.20$\pm0.15 \times$ solar. This is one of the highest C/O abundances at z$>$10, and it is consistent with either PopII and PopIII enrichment paths. Despite this source being extremely compact, with a half-light radius of 73$\pm$10 pc, we see no increased equivalent width of NIV] and NIII] emission lines as seen in some other compact sources at similar redshifts, a potential signature of second-generation stars in GCs. Overall, this galaxy exhibits low metallicity and high ionisation parameter consistent with intense star-formation or AGN activity in the early Universe, possibly observed before the enrichment by the second generation of stars in proto-globular clusters in the core of the galaxy.

\end{abstract}




\section{Introduction} 

With the launch of \jwst we are now able for the first time to study the physical properties of galaxies above z$>$9 using their rest-frame optical and UV emission
\citep[e.g. ][]{curtis-lake_2023, robertson_JOF_LF_2024, Bunker23gnz11, Maiolino23gnz11, Hsiao23, Arrabal_Haro_nature_2023, Castellano24, DEugenio24z12, carniani_z14_2024}.  These observations challenge our predictions of the high-redshift Universe based on the knowledge of the lower redshift universe, whether through increasing ISM density \citep{reddy_ism_2023}, ionisation parameter \citep{Cameron23}, temperature \citep{curti_smacs_2023}, star-formation burstiness ( \citealt{Endsley24, Looser23_SFH, Boyett24,dressler24})  or decreasing metallicity \citep{Schaerer22, curti_jades_mzr_2024, nakajima_mzr_ceers_2023}. Specifically, extrapolation of the expected trends to galaxies at z$>$9 do not fully explain the new population of galaxies, characterized by high UV luminosities \citep{Castellano24, carniani_z14_2024, Naidu2025}, metallicities \citep{Castellano24, Carniani2025}, and hints of settled disks \citep{Scholtz+2025}. 

Furthermore, observations of high-z galaxies have found evidence for peculiar chemical abundances.
For instance, while there is a large population of sources with decreasing carbon-to-oxygen abundance at low metallicities
\citep{arellano-corodva_2023, jones_CO_z6_2023, stiavelli+2023,Curti2025}, as expected by simple chemical evolutionary models \citep{maiolino_re_2019}, JWST studies have also found some galaxies that display enhanced C/O at low metallicity \citep{Castellano24, DEugenio24z12,Nakajima2025}, which may be connected to the chemical enrichment by the first generation of stars (PopIII).
Additionally, while most galaxies at low metallicities show decreased N/O abundance ratios, again in line with chemical evolutionary models, JWST studies have also found a population of galaxies with enhanced N/O
\citep{topping_z6_lens_2024, Isobe2025, Naidu2025, Cameron23gnz11, Scholtz_CO30_2025}. The origin of this enhancement is debated: it could be connected with the metal enrichment by the early population of AGB stars \citep{DAntona2023}, or Wolf-Rayet stars, possibly convolved with specific star formation histories \citep{kobayashi_ferrara_gnz11_2024, McClymont25_NO}, or very massive stars \citep{Vink2023,charbonnel_gnz11_2023}, enrichment by star formation that lacks feedback in the first few Myr \citep{Renzini2023},
or differential outflows \citep{Rizzuti2025}. Regardless of the origin of the N/O enhancement, it seems that this chemical enrichment pattern is connected with the early formation of proto-globular clusters. Indeed, \citep{Ji_2025_GC} found strong similarities with the chemical enrichment patterns of the second generation of stars in Globular Clusters (GCs). This is in line with the finding of a chemical and density stratification in these early galaxies, whereby the most nitrogen rich regions are associated with the densiest regions in the same galaxies \citep{ji_nitrogen_AGN_z5_2024, Ji_2025_GC, Hayes25}, and it is also in line with the finding that nitrogen rich galaxies tend to be very compact \citep{Schaerer24, topping_z6_lens_2024, Harikane25,Naidu2025}. The finding that many of the nitrogen rich galaxies host AGN \citep[e.g.][]{ji_nitrogen_AGN_z5_2024,Maiolino23gnz11,Napolitano2024} and, viceversa, the fact that most AGN found by JWST are nitrogen-rich \citep{Isobe2025}, also suggests a connection between black holes seeding and the early merging of proto-globular clusters in the cores of early galaxies \citep{Rantala2025}. 

Many of these findings, however, remain tentative and are restricted to the bright population of galaxies in the early Universe for which the detection of nebular lines is easier. It is important to explore the properties of fainter systems, which are more representative of the overall galaxy population, as well as to study more distant galaxies that may probe the earliest stages of star and black hole formation.
This is one of the main objectives of the Guaranteed Time Observing programme called JWST Advanced Deep Extragalactic Survey (JADES; \citealt{bunker_hst_deep_DR_2023, eisenstein_JOF_2023,rieke_jades_DR_2023, DEugenio24_DR3}).
In this work, we present deep PRISM and grating NIRSpec-microshutter array observations of \target originally identified through a Lyman-break in the NIRCam imaging in \citet{hainline_zgtr10_jades_2024} and observed with NIRSpec/MSA as part of JADES and presented as part of a larger sample of high-z galaxies in \citet{Tang2025b}.

In \S~\ref{sec:obs} we present our observations and data reduction, in \S~\ref{sec:analysis} we describe our emission line fitting and SED fitting. In \S~\ref{sec:results} we present our results and finally in \S~\ref{sec:summary} we summarise and discuss our results.
Throughout this work, we adopt a flat $\Lambda$CDM cosmology: H$_0$= 67.4 km s$^{-1}$ Mpc$^{-1}$, $\Omega_\mathrm{m}$ = 0.315, and $\Omega_\Lambda$ = 0.685 \citep{2020A&A...641A...6P}. We use vacuum wavelengths for the emission lines throughout the paper.

\section{Observations and data reduction}
\label{sec:obs}

\subsection{NIRSpec data}
\label{sec:nirspecdata}
The	NIRSpec data used in this work are part of the Guaranteed Time Observations program ID 1287 (PI: K. Isaak). The observations using the microshutter array (MSA; \citealt{jakobsen_nirspec_2022, Ferruit22}) mask were designed using eMPT software \citep{bonaventura_empt_2023} and proceeded using the same method as described in \citet{bunker_hst_deep_DR_2023} and \citet{DEugenio24_DR3}. During the design of the mask, the pointing was optimised for the highest priority sources in the catalogue, including \target. The procedure guarantees that all targets are located within $\sim90$ and $\sim220$ mas from the centre of the MSA shutter (with a size of 0.2”$\times$0.46”) in the dispersion and spatial directions, respectively. The spectra of high-priority targets (\target included) were protected against possible spectra overlap with other (lower priority) sources.

The target was observed in three visits in January 2024 (Visit 1+2) and January 2025 (Visit 3). The disperser-filter configurations employed in the programme were PRISM/CLEAR, G140M/F070LP, G235M/F170LP, G395M/F290LP, and G395H/F290LP. The first four spectral configurations provided spectroscopic data with spectral resolving power of ${\rm R=\Delta\lambda/\lambda\sim100}$ and ${\rm R\sim1000}$ in the wavelength range between 0.6~$\mu$m and 5.5~$\mu$m. This is done by performing a linear fitting of the S-FLATs and F-FLAT curves between 5.2 and 5.3$\mu$m and extending the profile up to 5.5$\mu$m. The G395H/F290LP disperser-filter configuration nominally covered the wavelength range 2.87–5.27~$\mu$m with the spectral resolving power of ${\rm R\sim2700}$, however, the wavelengths above 4.1 $\mu$m fall outside of the detector. In this work, we will focus our effort on the PRISM/CLEAR and R1000 observations. We applied a slit-loss correction appropriate for point sources. We note that this slit-loss correction is optimised for a 5-pixel aperture. In this work, we use the 3-pixel extraction due to the higher SNR of the emission lines in the data. However, we verified that fluxes measured from the 3-pix and 5-pix are consistent within 1$\sigma$. 

The PRISM observations were set up as four sequences of three nodding exposures at each visit, while one sequence of three nodded exposures was used for the spectral configuration of the gratings. Each nodded exposure sequence consisted of six integrations of 19 groups in NRSIRS2 readout mode \citep{Rauscher:2012},  resulting in total exposure time of 99788 and 24947 seconds in PRISM and per R1000 grating, respectively.

\begin{figure*}
        \centering
	\includegraphics[width=0.95\textwidth]{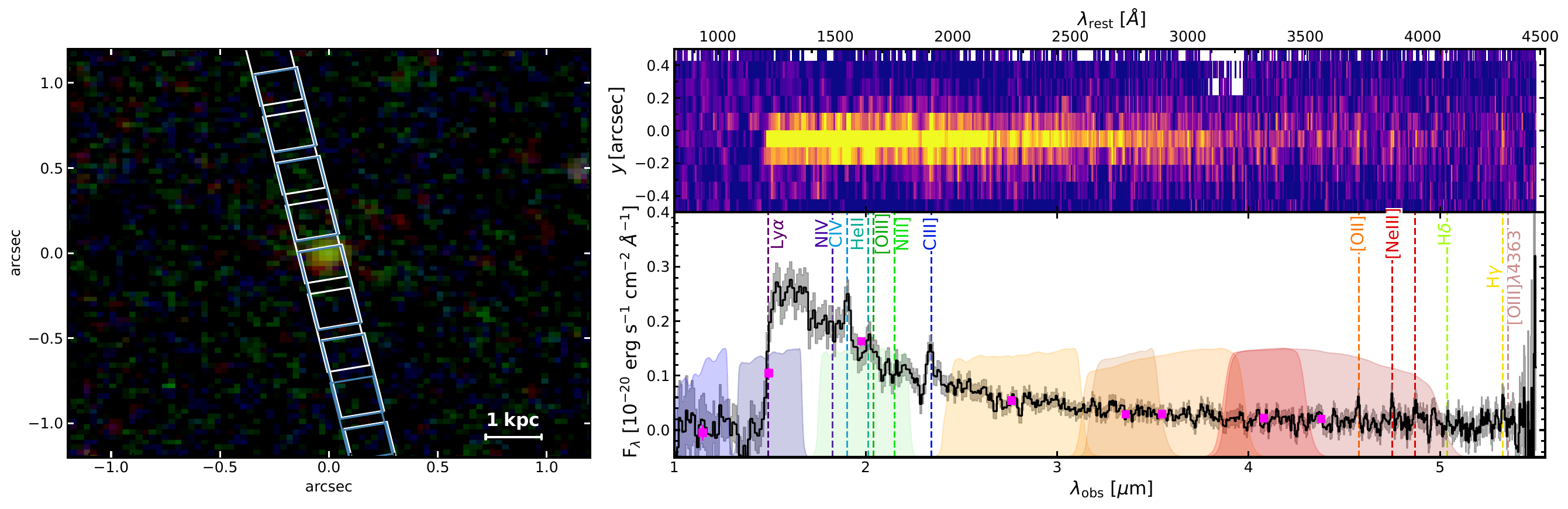}
    \includegraphics[width=0.95\textwidth]{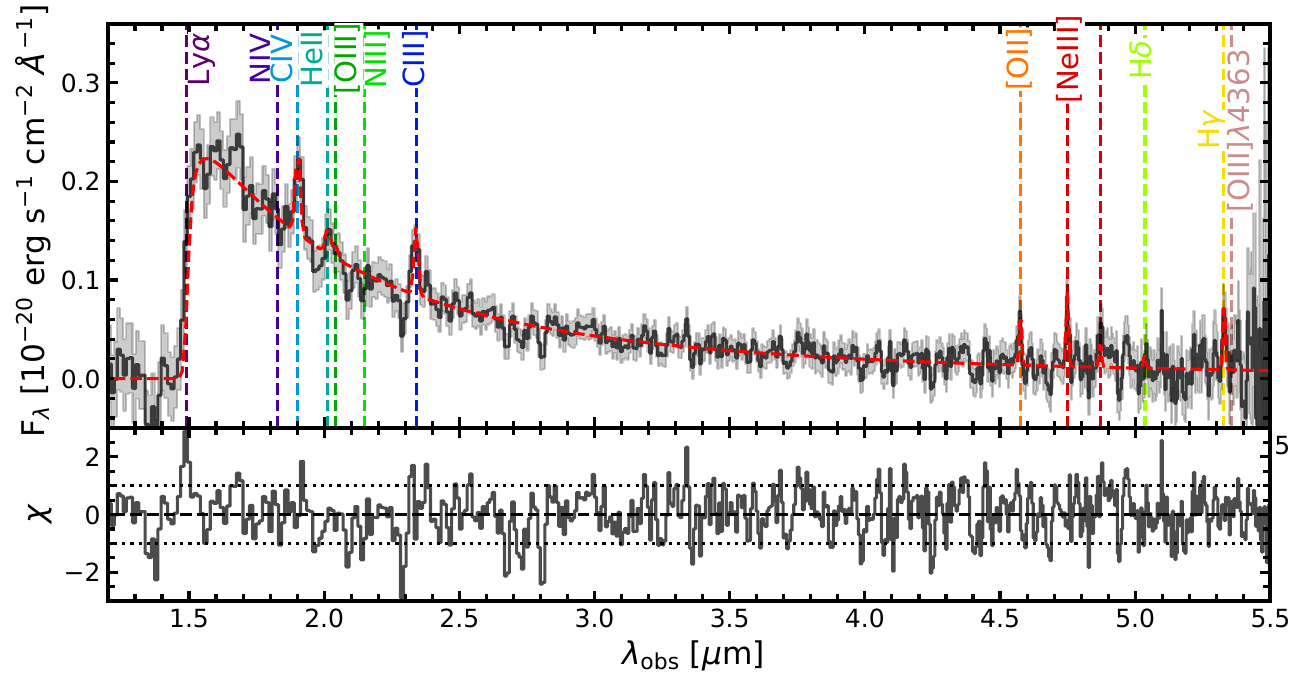}
    \caption{Top left: NIRCam RGB (F150W, F277W, F444W) image image with shutters from Obs 1+2 and Obs 3 overplotted as while and blue regions. We show the scale bar showing a size of 1 kpc in the bottom right corner. Top right: 1D (bottom panel) and \& 2D spectrum (top panel) of \target. The magenta squares show the extracted NIRCam photometry with the filters' transmission curves as coloured shaded regions. Bottom panel: Final 1D spectrum (black line) and its uncertainties (black shaded region) bootstrapped from the 72 exposures from our observations.  We show the location of the main emission line of interest as coloured vertical dashed lines. Bottom panels: NIRSpec/MSA PRISM spectrum of the rest-frame UV emission and the best-fit model. The data is shown as a black solid line with a shaded region showing the 1$\sigma$ uncertainties. The best-fit model is shown as a red dashed line. We highlight the emission lines of interest as dashed vertical lines. 
    }
    \label{fig:Spectrum}
\end{figure*}

We made use of the NIRSpec GTO pipeline \citep{JADES_DR4} to process the data. The pipeline was developed by the ESA NIRSpec Science Operations Team and the NIRSpec GTO Team. A general overview of the data processing is reported in \citep{JADES_DR4}. To optimise the signal-to-noise ratio of the data, we used the 1D spectra extracted from an aperture of 3 pixels, corresponding to 0.3~arcsec, located at the target position in the 2D spectra. Unlike the standard pipeline version, we exploited the additional data available on the detector that exists past the nominal 5.3 $\mu$m red cutoff, extrapolating the calibration files to 5.5 $\mu$m, which allows us to investigate the H$\gamma$ and \OIIIL[4363] emission lines that would not be present in the standard data reduction. More information about this procedure is covered in \citet{JADES_DR4}. We note that when combining the individual exposures, we scaled the Visit 3 exposures to match the PRISM to the NIRCam photometry. We describe the procedure in Appendix \ref{sec:obs_scaling}. We show the final combined PRISM spectrum of the 72 individual exposures in the right panel of Figure \ref{fig:Spectrum}, along with the NIRCam photometry and imaging.

\subsection{Imaging data}\label{sec:nircam}

The imaging data was obtained with The Near-Infrared Camera (NIRCam) on board from the James Webb Space Telescope (\JWST) by JADES \citep[PIDs: 1180, 1210, 1286][]{Eisenstein_JADES_2023, eisenstein_JOF_2023, rieke_jades_DR_2023} and JEMS \citep[PID: 1963][]{Williams23} and are processed through the JADES custom reduction pipeline.

The full description of the data reduction is in \citet{rieke_jades_DR_2023}. Here, we briefly summarise the procedure. The data was reduced with \JWST Calibration Pipeline version 1.9.6, incorporating the latest Calibration Reference Data including dark frames, distortion maps, bad pixel masks, read noise, superbias, and flat-field files.  At the detector level, the pipeline performs a group-by-group correction with dark subtraction, linearity calibration, and cosmic ray identification, followed by ramp fitting to recover count-rate images. Furthermore, the pipeline performs further calibration steps, including flat-fielding, photometric calibration, and background subtraction to ramp-fitted images. The exposures are aligned using a custom version of TweakReg to align exposures and to match the sources across images to calculate relative and absolute astrometric solutions. The calibrated, aligned exposures are then combined into mosaics using the \JWST Stage 3 pipeline.

For the photometric analysis presented in this work, we fit \target using ForcePho (Johnson B.
et al., in prep., but also \citealt{Robertson23,Tacchella23, Baker25disc}). We report the final fluxes and morphological parameters from the NIRCam imaging (using F090W, F115W, F150W, F200W, F277W, F335M, F356W, F410M and F444W filters) in Table \ref{tab:nircam}, with the full modelling results summarised in the Appendix Figure \ref{fig:forcepho}.

\begin{table}
    \caption{Results of the ForcePho fit to the NIRCam imaging of  \target.}
    \centering
    \begin{tabular}{lc}
        \hline
        \hline
        Filter Flux & \target\\
         & [nJy] \\
        \hline
        F090W & $0.12\pm 0.60$\\
        F115W & $-0.21\pm 0.62$ \\
        F150W & $ 7.8\pm 0.75$  \\
        F200W & $21.28\pm 1.02$  \\
        F277W & $13.71\pm 0.68$ \\
        F335M & $10.82\pm 1.8$ \\
        F356W & $12.36\pm 0.73$  \\
        F410M & $12.24\pm 1.45$  \\
        F444W & $13.11\pm 0.99$  \\
        \hline
        \hline
        Morphological information \\
        \hline
        PA/deg &  -31.11988 \\
        r$_{\rm half}$/arcsec &  0.018 $\pm $0.0025\\
        r$_{\rm half}$/pc &  73 $\pm $10\\
        n$_{\rm sersic}$ & 2.13 $\pm$ 0.86 \\
        \hline
    \end{tabular}
    \label{tab:nircam}
\end{table}

\section{Data Analysis}\label{sec:analysis}

\subsection{Emission line fitting}\label{sec:eml_fit}

We performed the emission line fitting in the PRISM and in the R1000 using multiple separate techniques to verify the detection of emission lines. Formally, the error on each parameter of the fit is evaluated by exploiting the output pipeline error spectrum. However, we also perform additional tests to assess the robustness of low-significance detections in the PRISM spectrum by leveraging the 72 individual exposures in the PRISM observations.

\begin{figure*}
        \centering
	\includegraphics[width=0.8\paperwidth]{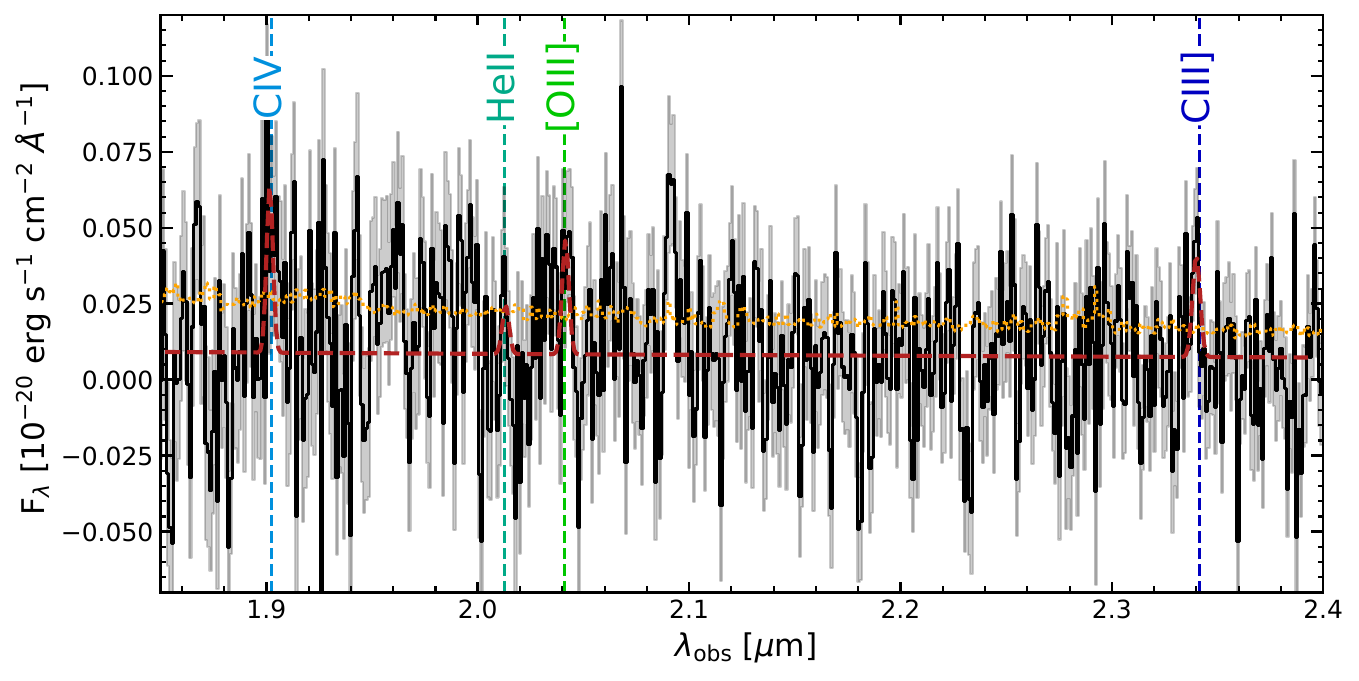}
    \caption{G235M F170LP spectrum and the best fit covering the rest-frame UV emission lines: \CIVall, \HeIIL[1640], \semiforbiddenEL[O][\textsc{iii}][1663] and \CIIIall. We report a 3.2$\sigma$ detection of \CIIIall emission line doublet. 
    }\label{fig:R1000}
\end{figure*}

The standard fitting was performed using \texttt{QubeSpec}'s fitting routine \footnote{\url{https://github.com/honzascholtz/Qubespec}} \citep{Scholtz_COS30_2025}. We fitted the \NIVall, \NIIIall, \CIVall, \CIIIall, OIII]$\lambda$1663 and \HeIIL[1640], \OIIall, \NeIIIL[3869], H$\delta$, H$\gamma$ and \OIIIL[4363] emission lines along with the continuum. The continuum was modelled as a single power-law assuming mean-density IGM with a neutral fraction of x$_{HI}$ = 1 at the redshift of the galaxy (see \citealt{Witstok+2025} for more details), while each emission line was modelled as a Gaussian profile. As the emission lines are unresolved in the PRISM observations, we set the intrinsic FWHM of each emission line to an arbitrary 100 \kms, which is then convolved with the line spread function (LSF) of the PRISM at the wavelength of the emission line. The LSFs are estimated for a point source as described in \citet{de_graaff_jades_2024}. We fit the spectrum between 1 and 5.5 $\mu$m, corresponding to 1000 and 4400 $\AA$ in the rest-frame. We show the best-fit to the spectrum in Figure \ref{fig:Spectrum} and we report the measured fluxes, equivalent widths and SNR in Table \ref{tab:fluxes}. For the R1000 observations, we performed similar fitting, except we set the FWHM of the lines as a free parameter. We show the best fit of the \CIVall, \CIIIall, \HeIIL[1640], \CIIIall and \OIIIL[1663] emission lines in Figure \ref{fig:R1000}.

We also estimate the fluxes and the SNR of the emission lines using bootstrapping of the individual 72 exposures. We generated 5000 bootstrapped spectra by randomly sampling (with replacement) over the set of individual sub-spectra, after four passes of iterative 3$\sigma$-clipping to remove any remaining outliers not flagged by the pipeline. For each iteration of the bootstrapped spectrum, we fitted the model described above to estimate the continuum shape and measured fluxes. The final measured fluxes of the emission lines from the bootstrapping approach is the mean of the distribution, with error estimated as the standard deviation. These bootstrapped uncertainties are considered to be more conservative than the flux density uncertainty estimates from the pipeline as they natively consider all sources of noise, including correlated noise from resampling the NIRSpec spectra (see also \citealt{maseda_xmps_lae_2023, hainline_zgtr10_jades_2024, DEugenio24_DR3, Curti24_9.4}). A more detailed discussion of the noise model in NIRSpec spectra will be presented in a forthcoming paper (Jakobsen et al., in prep). We report the measured fluxes, equivalent widths and SNR in Table \ref{tab:fluxes}.

Overall, the two methods give consistent SNR and measured fluxes with uncertainties. Although the bootstrapping technique is supposed to be more conservative compared to the MCMC fitting. However, this shows that the scaling GTO pipeline error for the PRISM spectrum being intrinsically is sufficient to correctly account for the correlation induced by the spectral resampling, as noted by \citet{Curti24_9.4}.

\begin{table*}
    \caption{Emission line fluxes and equivalent widths in \target in PRISM \& R1000 observations.}
    \centering
    \begin{tabular}{lccc|ccc|ccc}
        \hline
        \hline
        & \multicolumn{3}{c}{PRISM MCMC} &  \multicolumn{3}{c}{PRISM Bootstrapping} & \multicolumn{3}{c}{R1000} \\
        Emission line & Flux & EW & SNR & Flux & EW & SNR &  Flux & SNR & FWHM \\
         & 10$^{-19}$ erg s$^{-1}$ cm$^{-2}$ & $\AA$ & & 10$^{-19}$ erg s$^{-1}$ cm$^{-2}$ &  $\AA$ & & 10$^{-19}$ erg s$^{-1}$ cm$^{-2}$ &  & \kms \\
        \hline
        \semiforbiddenEL[N][\textsc{iv}][\textlambda\textlambda][1483,][86]      & $<1.4$ & $<2.3$ & -   & $<1.6$  & $<2.5$ &  - & $<2.7$ & -& -  \\
        \CIVall      & 2.21$^{+0.38}_{-0.44}$ & 6.7$^{+1.4}_{-1.3}$ & 5.7   & 2.25$^{+0.71}_{-0.51}$  & 6.7$^{+1.7}_{-1.5}$ &  4.26 & $<2.4$ & -& -  \\
        \HeIIL[1640] & $<1.11$ &$<4.5$ & -   & $<1.32$  & $<4.5$ &  - & $<1.05$ &  - & - \\
        \semiforbiddenEL[O][\textsc{iii}][\textlambda\textlambda][1661,][66] & $<0.96$ & $<2.7$ & -   &    $<1.32$ & $<3.5$ & - & $<1.6$ &   - &- \\
        \semiforbiddenEL[N][\textsc{iii}][\textlambda\textlambda][1748-][54]      & $<0.9$ & $<3.6$ & -  & $<1.1$ & $<3.7$ & - & $<1.5$ & -& -  \\

        \CIIIall     & 1.56$^{+0.27}_{-0.25}$ & 8.6$^{+1.5}_{-1.4}$ & 5.9   & 1.23$^{+0.51}_{-0.27}$  & 8.1$^{+1.6}_{-1.7}$ & 4.7 & 1.37$^{+0.47}_{-0.43}$ &3.13   & 520$_{-170}^{+190}$\\
        \hline
        \OIIall       & 0.56$^{+0.15}_{-0.15}$ & 20.9$^{+5.7}_{-6.1}$& 3.8 & 0.57$^{+0.22}_{-0.16}$ & 21.0$^{+7.1}_{-7.2}$ & 3.0 & $<0.72$ &  - & - \\
        \NeIIIL[3869] & 0.75$^{+0.14}_{-0.15}$ & 32.6$^{+6.2}_{-6.1}$& 5.2 & 0.81$^{+0.25}_{-0.16}$ & 33.5$^{+7.2}_{-7.2}$ & 5.06 & $<1.5$ &  - & -\\
        H$\delta$ & $<0.42$ & $<16.5$ & - & $<0.33$ & $<17.5$ & -  & $<2.4$ &  - & -\\
        H$\gamma$ & 0.75$^{+0.21}_{-0.19}$ & 43.0$^{+10}_{-12}$ & 3.9 & 0.73$^{+0.22}_{-0.20}$ &43.0$^{+56}_{-30}$ &  3.68 & $<2.3$ & - &- \\
        \OIIIL[4363] & $<0.63$ & $<36.0$ & - & $<0.72$ & $<44.2$ &  - & $<2.4$ & - &- \\
        \hline
    \end{tabular}
    \label{tab:fluxes}
\end{table*}

\subsection{\normalfont\scshape{beagle} \normalfont{and} \normalfont\scshape{beagle-agn} \normalfont{SED fitting}}\label{sec:beagleSED}

In order to derive the physical properties of \target we fit the PRISM spectrum using \beagleagn\ \citep{Vidal-Garcia24}, an extension to the Bayesian SED fitting code \beagle\ \citep{Chevallard2016} which incorporates a prescription for the narrow line region (NLR) component \citep{Feltre16}. We mask the region around Ly-$\alpha$ as our modelling does not properly account for the (resonant) radiative transfer of Ly-$\alpha$ photons, as well as \NeIII\ due to its known inconsistency with model coverage in many high redshift galaxies (\citeauthor{Silcock24} \citeyear{Silcock24}, see also \citeauthor{Shapley_2025_AURORA} \citeyear{Shapley_2025_AURORA})
. Owing to the blue $\beta$ slope of \target\ (see Section \ref{sec:escape}), we allow the Lyman-continuum photon escape fraction f$\mathrm{_{esc}}$ to vary between [0.0 - 1.0] assuming a picket-fence model. We assume a \citet{Chabrier03} initial mass function (IMF) with an upper mass cut-off of 300 M$_{\odot}$. The star-formation history (SFH) is described as a delayed exponential function with the star formation rate as a function of time described by $\mathrm{SFR}(t)\propto t\,\mathrm{exp}(-t/\tau_\mathrm{SFR})$ where $t$ is the age and \tausfr is the characteristic timescale (both in yr).
An additional burst component of 10 Myr, during which the star-formation rate is constant, constitutes the recent star formation history and is the period over which the SFR 
is calculated. The maximum age of the stars in the galaxy is denoted by $\mathrm{log(t_{max}/yr)}$. This overall approach attributes the strength of the (line and continuum) nebular emission to the recent $10$ Myr SFH. The stellar mass $\log(\mathrm{M_{\star}}/\Msun)$ includes mass formed during both the extended and recent star formation periods whilst accounting for the return fraction of stellar mass to the ISM as the stars evolve. The \textit{total} stellar mass formed, not accounting for the return fraction, is denoted $\log(\MtotInLog/\Msun)$.

The parametrisation of the star-forming nebulae uses the photoionisation model presented in \cite{gutkin_modelling_2016}. 
We adopt an effective galaxy-wide metallicity, log($\Zism/\Zsun$), and ionization parameter, \logUs, for the \HII regions surrounding young stars. The metallicity of stars is set equal to log($\Zism/\Zsun$), in the absence of features in the continuum able to independently constrain stellar metallicity.  
The interstellar metallicity log($\Zism/\Zsun$) is linked to the gas-phase form via the dust-to-metal mass ratio parameter $\xid$, which we fix to 0.1, and accounts for the differential depletion of heavy elements onto dust grains. To account for dust attenuation, we employ the \citet{charlot00} two-component model and fix the fractional attenuation due to the diffuse ISM to $\mu=0.4$. The dust attenuation is parametrised by the effective $V$-band attenuation optical depth, $\tauV$. We also account for a varying nebular carbon-to-oxygen abundance $\mathrm{(C/O)^{\HII}/(C/O)^{\odot}}$ in the range [0.1 – 1.4], where we assume a solar carbon-to-oxygen abundance value of $\mathrm{(C/O)^{\odot} = 0.44}$.

With \beagleagn, we can remove the NLR component such that a galaxy with only a star-forming component is modelled (as described so far), or include the NLR component should the galaxy be considered to host an obscured AGN. In the latter case, both \HII\ and narrow-line region-associated parameters are constrained simultaneously. Given the ambiguity in the dominating ionising source for \target, as expanded in Section \ref{sec:Diagnostics}, we consider both cases in our SED fitting. We perform one fit which includes both an SF and NLR component (named \ticks{SF+AGN}), and a version of this which omits the NLR component (named \ticks{SF-only}). 

For our \ticks{SF+AGN} fit, we characterise the properties of the NLR using the models of \cite{Feltre16}, themselves updated in \cite{Mignoli19} to include better parametrizations for the NLR inner-radius and internal microturbulence of the narrow line-emitting gas clouds. In this model, the thermal emission from the AGN accretion disk is characterised by a broken power law, with the slope between $0.001\leq\lambda/\mu\mathrm{m}\leq 0.25$  parametrized by \PLalpha. This can vary between $-2.0 < \PLalpha < -1.2$ though here we fix it to $\PLalpha = -1.7$ following the suggestions within \cite{Vidal-Garcia24}. The integral under the broken power law provides an estimate of the accretion disk luminosity, \Lacc, a proxy for the bolometric luminosity. Additional NLR-related parameters corresponding to the \cite{Feltre16} models are NLR metallicity $\mathrm{\ZAGN/\Zsun}$ and NLR ionising parameter \logUsAGN. Akin to the ISM metallicity, the NLR metallicity is related to its gas-phase form via the NLR dust-to-metal mass ratio \xidAGN, which we fix to 0.1. Within the framework of NLR and nebulae emission, the light emitted from the NLR is attenuated by dust within the NLR itself (accounted for in the \citealp{Feltre16} models) as well as dust in the diffuse ISM. Additionally to varying the nebular carbon-to-oxygen abundance, we vary the NLR carbon-to-oxygen abundance $\mathrm{(C/O)^{NLR}/(C/O)^{\odot}}$ in the same range [0.1 - 1.4]. Allowing variations of both nebular and NLR carbon-to-oxygen abundances in our fitting facilitates a test to see what abundances are required to reproduce the strength of the carbon UV lines in \target. The parameters described here for both the \ticks{SF+AGN} and \ticks{SF-only} fits, and their corresponding priors,  are presented in Table \ref{tab:fiducialparameters}. Besides the accretion disk luminosity $\Lacc$ and power-law slope $\PLalpha$, parameters in Tab. \ref{tab:fiducialparameters} that are unique to the \ticks{SF+AGN} fit (in addition to the remaining parameters) are indicated with a \ticks{NLR} superscript.

\begin{table}
    \centering
    \caption{Prior limits, fixed values and other parameters used in this work's \beagle\ and \beagleagn\ fits to \target. Priors described with $\mathcal{N}{[\mu,\sigma^{2}]}$ notation denote a Gaussian profile with mean $\mu$ and standard deviation $\sigma$.}
    \begin{tabular}{c c}
    \hline
    \hline
       Parameter & Prior\\
       \midrule
        
        $\tauV$ & Exponential $\in [0,5]$ \\

        $\log(\Z/\Zsun)$ & Uniform $\in [-2.2, 0.4]$\\

        $\logUsHII$ & Uniform $\in [-4, -1]$\\

        \sfrconsistent\ & Uniform $\in [-4, 4]$\\
        
        $\log(\tausfr/\yr)$ & Uniform $\in [6,12]$\\
        
        $\log(\ZAGN / \Zsun)$ & Uniform $\in [-2,0.3]$\\

        $\mathrm{\logUsAGN}$ & Uniform $\in [-4,-1]$\\

        $\mathrm{log(\Lacc/ erg s^{-1})}$ & Uniform $\in [43, 48]$\\

        $z$ & $\mathcal{N}[11.27,0.02^{2}]$,  $\in [0,20]$\\

        $\log(\MtotInLog/\Msun)$ & Uniform $\in [6,12]$\\

        $\mathrm{f_{esc}}$ & Uniform $\in  [0.0,1.0]$\\
        
        $\mu$ & Fixed to 0.4\\
        
        $\xidAGN$ & Fixed to 0.1\\
        
        $\xid$ & Fixed to 0.1\\
        
        $m_{\textrm{up}}$/$\mathrm{M_{\odot}}$ & Fixed to 300\\

        $\frac{\mathrm{(C/O)^{HII}}}{\mathrm{(C/O)^{\odot}}}$ & Uniform $\in [0.1,1.4]$\\

        $\PLalpha$ & Fixed to -1.7\\

        $\frac{\mathrm{(C/O)^{NLR}}}{\mathrm{(C/O)^{\odot}}}$ & Uniform $\in [0.1,1.4]$\\        
        
    \bottomrule
    \end{tabular}
    
    \label{tab:fiducialparameters}
\end{table}

\begin{table}
    \centering
    \caption{Summary of the emission ratio line analysis (see \S \ref{sec:CO_abud}) and of \beagle\ and \beagleagn\ fits to \target. \ticks{SF $+$ AGN} is the NLR-inclusive fit and \ticks{SF-only} is the version which omits an AGN component.}
    \begin{tabular}{c c c}
    \hline
    Parameter & \multicolumn{2}{c|}{Value}\\
    \hline
    R.A. & \multicolumn{2}{c|}{53.11762}\\
    Dec & \multicolumn{2}{c|}{-27.88817}\\
       M$_{\rm UV}$ &\multicolumn{2}{c|}{-19.4}\\
       $\beta$ & \multicolumn{2}{c|}{-2.8$\pm$0.1} \\ 
       \hline
       log(SFR/\Msun yr$^{-1}$) (H$\gamma$) &\multicolumn{2}{c|}{0.50$\pm$0.15}  \\
       12+log(O/H) & \multicolumn{2}{c|}{7.75$\pm$0.30} \\
       log(C/O) & \multicolumn{2}{c|}{$>$-0.6} \\
       log U & \multicolumn{2}{c|}{-1.85$\pm$0.15} \\
       log(M$_{\rm dyn}$/\Msun) & \multicolumn{2}{c|}{9.0$\pm$0.5} \\ 
       \hline
       \beagle\ \& \beagleagn & SF $+$ AGN & SF-only\\
       \midrule
        $\tauV$ & $0.12^{+0.08}_{-0.08}$ & $0.05^{+0.03}_{-0.02}$ \\

        $\log(\Zhii/\Zsun)$ & $-1.66^{+0.28}_{-0.30}$ & $-1.55^{+0.12}_{-0.13}$\\

        12 + log(O/H)$^{\mathrm{HII}}$ &  $7.14^{+0.28}_{-0.30}$ & $7.23^{+0.13}_{-0.13}$\\

        $\logUsHII$ & $-2.08^{+0.81}_{-0.97}$ & $-1.55^{+0.25}_{-0.26}$\\

        \sfrconsistent\ & $-0.09^{+0.36}_{-0.13}$ & $0.32^{+0.11}_{-0.10}$\\

        $\mathrm{log(t_{max}/yr)}$ & $7.50^{+0.32}_{-0.32}$ & $7.37^{+0.27}_{-0.26}$\\

        $\log(\mathrm{M_{\star}}/\Msun)$ & $8.03^{+0.19}_{-0.18}$ & $7.80^{+0.12}_{-0.11}$\\

        $\log(\MtotInLog/\Msun)$ & $8.09^{+0.20}_{-0.19}$ & $7.84^{+0.13}_{-0.11}$\\

        $\mathrm{f_{esc}}$ & $0.30^{+0.21}_{-0.21}$ & $0.26^{+0.18}_{-0.18}$\\

        $\frac{\mathrm{(C/O)^{HII}}}{\mathrm{(C/O)^{\odot}}}$ & $0.95^{+0.36}_{-0.13}$ & $1.20^{+0.15}_{-0.15}$\\

        $\log(\ZAGN / \Zsun)$ & $-0.73^{+0.37}_{-0.39}$ & $-$\\

        12 + log(O/H)$^{\mathrm{NLR}}$ &  $7.92^{+0.39}_{-0.41}$ & $-$\\

        $\mathrm{\logUsAGN}$ & $-2.12^{+0.39}_{-0.36}$ & $-$\\

        $\mathrm{log(\Lacc/ erg s^{-1})}$ & $43.77^{+0.20}_{-0.19}$ & $-$\\

        $\frac{\mathrm{(C/O)^{NLR}}}{\mathrm{(C/O)^{\odot}}}$ & $1.12^{+0.22}_{-0.23}$ & $-$\\
        
    \bottomrule
    \end{tabular}
    
    \label{tab:beagleagnresults}
\end{table}

\section{Results}\label{sec:results}

In this section, we discuss the results based on the new NIRSpec observations. We present the redshift and the detection of rest-frame UV and optical emission lines \S~\ref{sec:eml_det}, we investigate the source of ionisation of the ISM in \S~\ref{sec:Diagnostics} and present SED modelling in \S~\ref{sec:beagleagnresults}. In \S~\ref{sec:ISM_prop} \& \ref{sec:CO_abud} we present the ISM and carbon abundances and, finally, in \S~\ref{sec:Mdyn} we present estimates of the dynamical mass of \target.

\subsection{Overview of the detected lines}\label{sec:eml_det}

We searched for emission lines in both PRISM and R1000 observations, and we fitted these lines as described in \S~\ref{sec:eml_fit}. We also verified the significance of these detections by bootstrapping the individual PRISM exposures as described in \S \ref{sec:eml_fit}. 

We detected the UV continuum and emission lines using NIRSpec/MSA PRISM spectroscopy, reliably detecting \CIVall, \CIIIall and \NeIIIL[3869] at $>4\sigma$ and \OIIall at $>3\sigma$ in the PRISM. With these emission lines, we confirmed the spectroscopic redshift of this source to z$_{\rm spec}$ = 11.275$\pm$0.003. We show the best fit of the emission lines and the continuum model in Figure \ref{fig:Spectrum}.

Furthermore, we detected \CIIIall in the R1000 observations at 3.1$\sigma$ significance, with a redshift of 11.272$\pm$0.0028, consistent with redshift from PRISM observations. The flux of \CIIIall in R1000 is consistent with the flux estimated from the PRISM observations, within 1$\sigma$. The rest of the emission lines detected in the PRISM are undetected in the R1000 observations, with the upper limits on the fluxes from R1000 consistent with measured fluxes in the PRISM. We summarised the measured fluxes in Table \ref{tab:fluxes}. From now on, we will use the emission line fluxes from the PRISM observations for our analysis.

We note that there is an excess of emission around Ly-$\alpha$ when modelling the continuum and emission lines using power-law + IGM transmission and emission lines (see bottom panel of Figure \ref{fig:Spectrum}), with one channel being over 5$\sigma$. This excess can be caused by inaccurate modelling of the line spread function (LSF) that we used to convolve the modelled spectrum or through complex radiative transfer around the Ly-$\alpha$ break. We see an excess of emission around Ly-$\alpha$ at 2.7$\sigma$ significance in R1000 G140M observations; however, the excess resulting Ly-$\alpha$ seen in R1000 would be easily detectable in the PRISM observations at >10$\sigma$. unless the FWHM of the Ly-$\alpha$ is broadened $>600$\kms. Therefore, these low significance residual features in the PRISM spectrum are most likely noise rather than actual detected emission. 

Furthermore, the two sets of observations performed in January 2024 have an emission peak near the \HeIIL[1640] at 2.7$\sigma$. However, this has not been confirmed by the additional observations in January 2025, and as such, we report only an upper limit on the \HeIIL[1640] emission line. Future deeper observations are required to further investigate the \OIIIL[1663] and \HeIIL[1640] emission lines. 

\subsection{Ionisation source}\label{sec:Diagnostics}

Given the high significance of the emission lines detected in our spectroscopic observations, we can investigate the source of ionisation in this galaxy: star-formation, AGN, or even PopIII stars. The ratios and equivalent widths of the detected emission lines (either collisionally excited or produced by recombination) can be modelled to determine the nature of the photoionisation source in this galaxy. To identify the source of photoionisation in \target, we rely on models from the literature: specifically, models from \citet{gutkin_modelling_2016}, \citet{Feltre16}, and \citet{Nakajima22}. The equivalent widths and fluxes of the \CIVall, \CIIIall and \HeIIL[1640] emission lines used in theare given in Table \ref{tab:fluxes}.

\begin{figure}
        \centering
        \includegraphics[width=0.99\columnwidth]{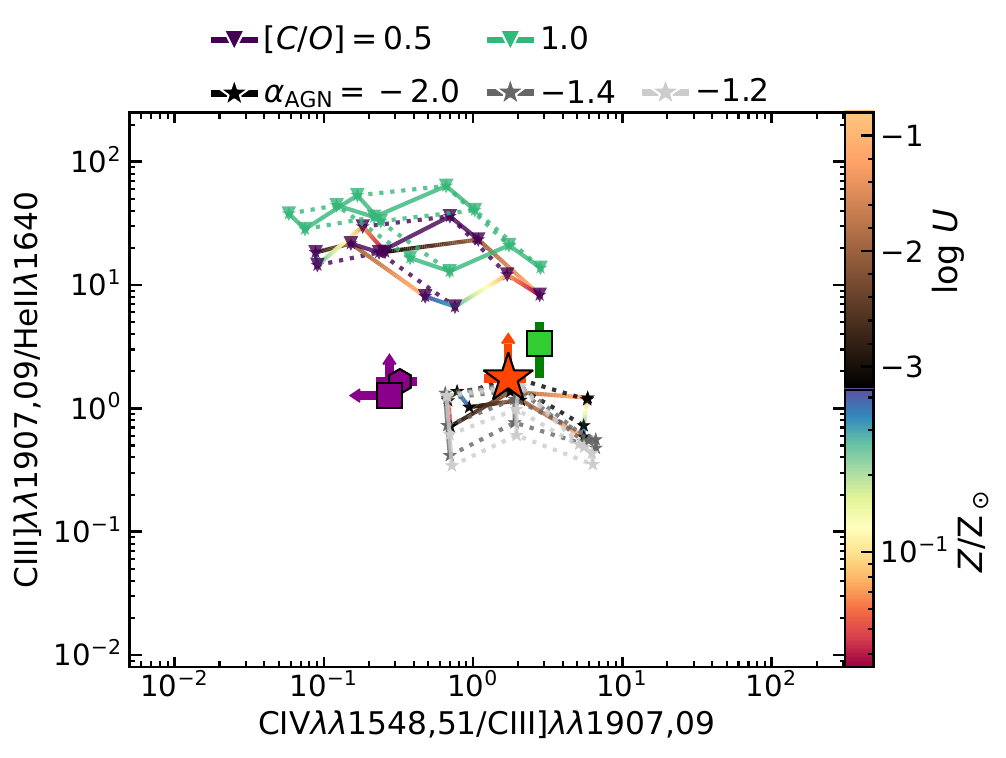}
        \includegraphics[width=0.99\columnwidth]{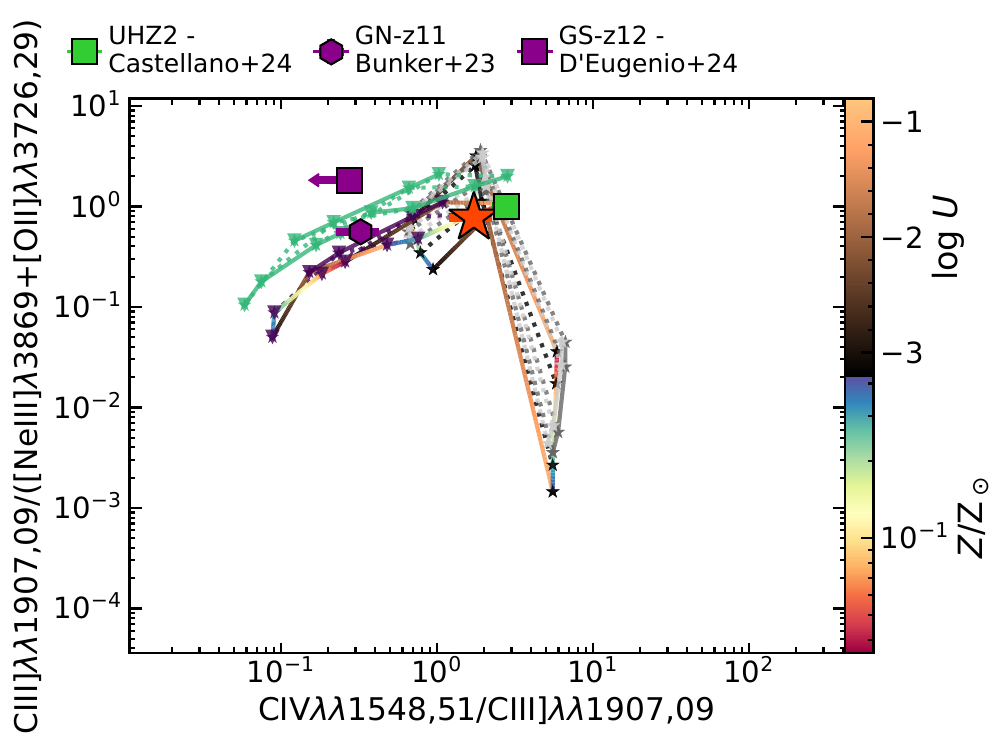}
        \includegraphics[width=0.99\columnwidth]
        {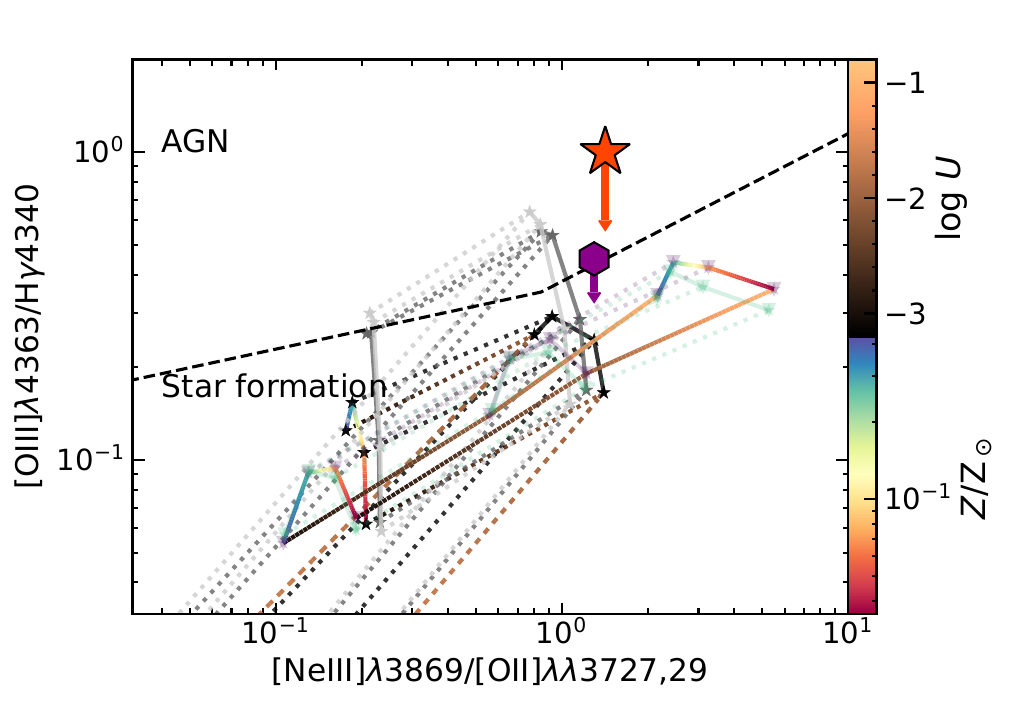}
    \caption{ UV and optical emission line diagnostics for \target (red star). Top panel: \CIIIall/\HeIIL[1640] vs \CIVall/\CIIIall . Middle panel: \CIIIall/(\NeIIIL[3869]+\OIIall) vs \CIVall/\CIIIall. Bottom panel: \OIIIL[4363]/H$\gamma$ vs \NeIIIL[3869]/\OIIall from \citet{mazzolari_new_2024, DEugenio24z12}. The grid on the diagnostics diagrams is from \citet{Feltre16} (AGN) \& \citet{gutkin_modelling_2016} (SF) for varying C/O abundances (SF) and accretion disk slopes (AGN). We show the emission line ratios of UHZ2 \citep{Castellano24} and GN-z11 \citep{Bunker23gnz11} as green and purple points.}
    \label{fig:UV_dig}
\end{figure}

The high equivalent width of the \CIVall emission line rules out sources requiring pristine gas such as Population III stars or direct collapse black holes in \citet[][]{Nakajima22}, which only occur in extremely low metallicity environments. The various line ratios (in particular the upper limit on HeII/H$\gamma$ and lower limit on CIII]/HeII) also exclude the scenario of self-polluted PopIII recently proposed by \citet{Rusta2025}. Therefore, here we focus on comparing our observations to the photo-ionisation models and observations of star-forming and AGN host galaxies. In Figure~\ref{fig:UV_dig}\footnote{The diagnostics are plotted using \url{https://github.com/fdeugenio/photoion_plot} }, we investigate the following line ratio diagnostics: \CIIIall/\HeIIL[1640] vs \CIVall/\CIIIall, \CIIIall/(\NeIIIL[3869]+\OIIall) vs \CIVall/\CII and \OIIIL[4363]/H$\gamma$ vs \NeIIIL[3869]/\OIIall. These line ratios are compared with models with different C/O abundances for star-formation models and AGN accretion disk slopes. In these diagnostics the emission line ratios observed in GS-z11-1 are consistent with AGN grids from \citet{Feltre16} as well as star-forming grids with high C/O of $>$50\% solar (log(C/O) = -0.65).

\begin{figure}
        \centering
        \includegraphics[width=0.99\columnwidth]{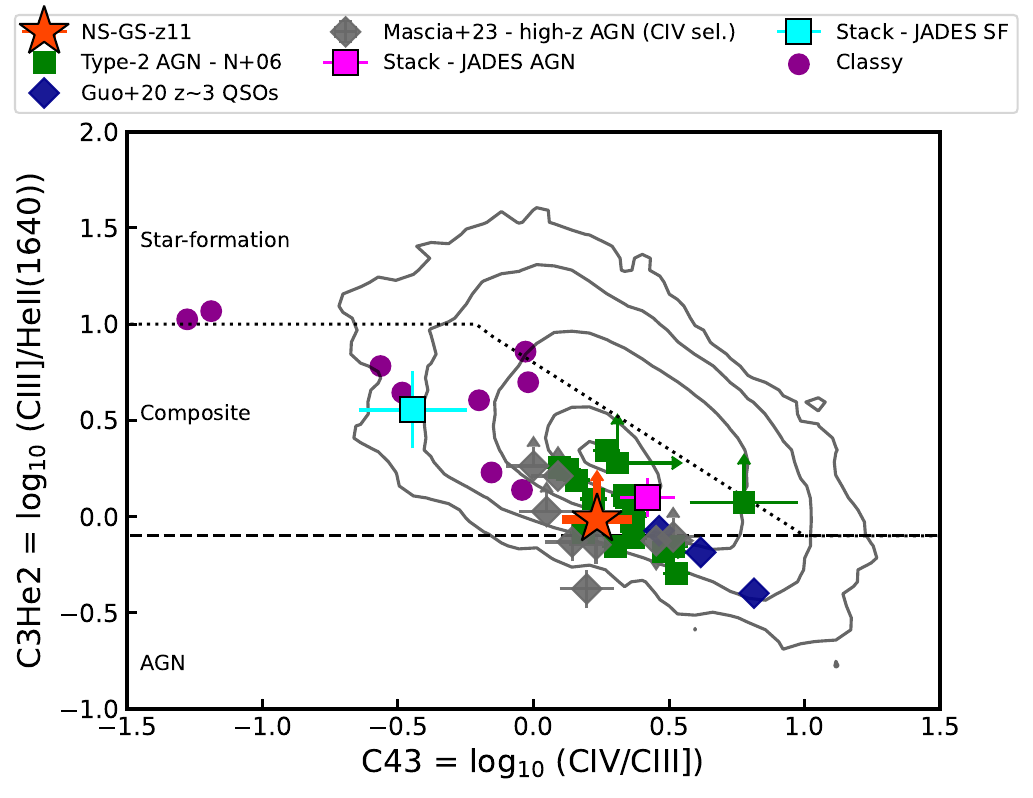}
    \caption{ UV emission line diagnostics for \target (red star): \CIIIall/\HeIIL[1640] vs \CIVall/\CIIIall and comparison to other objects and observations. The green squares, grey diamonds show the type-2 AGN \citep{Nagao06}
    and all \CIVall\space detections from \citet{Mascia23} as grey colour with squares and circles representing SF and AGN, respectively. The dark magenta circles show the local analogues of high-redshift galaxies \citep[from CLASSY survey;][]{Berg22}. The black contours show the narrow line emission of the SDSS quasars \citep{Guo_2020_SDSS}. The blue and light purple squares show the stacks of SF and AGN, respectively, from the JADES survey \citep{Scholtz23AGN}.   
    We also show the black dashed and dotted demarcation lines between AGN, star-forming galaxies and composite line ratios from \citet{Hirschmann22}.
    }
    \label{fig:UV_dig_Comp}
\end{figure}

In Figure \ref{fig:UV_dig_Comp} we show again the \CIIIall/\HeIIL[1640] vs \CIVall/\CIIIall where we now compare \target to quasars and Type-2 AGN from the literature \citep[][]{Nagao06, guo_sdss-iv_2018, Mascia23, tacchella_GS9422_2024}, SDSS quasars at z$\sim$ \citep{Guo_2020_SDSS}, star-forming galaxies from CLASSY \citep{Mingozzi23} and stacked UV spectra of star-forming and AGN host galaxies from JADES (\citealt{Scholtz23AGN}). 
We also draw with dotted lines the demarcation between AGN, star forming and composite galaxies according to 
\citet{Hirschmann_UV_diags_2019}. Overall \target appears primarily consistent with AGN observed in the literature and in the composite region identified by \citet{Hirschmann_UV_diags_2019}.
However, since the \HeIIL[1640] is not detected, our object can also lie in the star-forming part of the diagram.

\begin{figure}
        \centering
	\includegraphics[width=0.99\columnwidth]{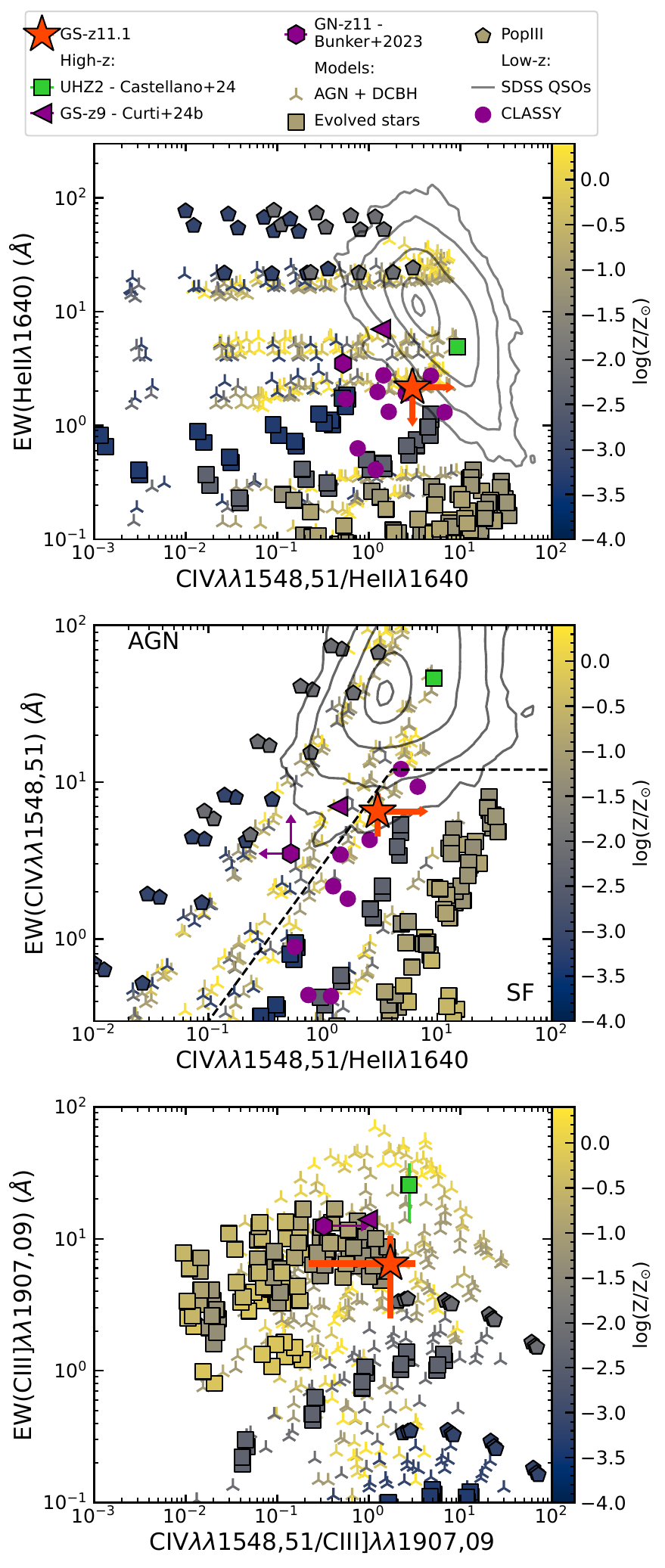}
    \caption{AGN, PopIII and evolved stars (PopI and PopII) diagnostics from \citet{Nakajima22, DEugenio24z12}. Top: EW of \HeIIL[1640] vs \CIVall/\HeIIL[1640]. Middle panel: EW of \CIVall vs \CIVall/\HeIIL[1640]. Bottom panel: EW of \CIIIall vs \CIVall/\CIIIall. \target is marked as a red star, while we show AGN candidate UHZ2 \citep{Castellano24} as dark green square. The purple points show local dwarf analogues from CLASSY \citep{Mingozzi23}}
    \label{fig:EW_dig}
\end{figure}

We measured the rest-frame equivalent width (EW) of the \CIVall, \CIIIall and \HeIIL[1640] as 6.7$\pm$1.4 $\AA$ and 8.6$\pm$1.3 $\AA$ and $<2.5$ $\AA$, respectively. We plot the EW of these lines vs UV emission line ratios in Figure \ref{fig:EW_dig}. We compared the observed EW of \CIVall and \HeIIL[1640] with photo-ionisation models from \citet{Nakajima22} as well as local high-z analogues from \citet{Mingozzi23} and high-z galaxy UHZ-2 \citep{Castellano24}. Similarly to the analysis of the emission line ratios, \target is in the region of these diagnostics diagrams consistent with both star-formation and AGN ionisation. Overall, based on the slew of emission line ratio and equivalent width diagnostics diagrams \target is consistent with both star-formation and AGN ionisation.

\subsection{$\beta$-slope and escape fraction}\label{sec:escape}

From the visual inspection of the spectrum (see Figure \ref{fig:Spectrum}), it is clear that \target is characterized by a very steep $\beta_{\rm UV}$ slope. We fitted the $\beta_{\rm UV}$ slope using multiple different procedures to verify its steepness, with either selecting the emission line free windows described in detail in \citet{saxena24}, which are based on the \citet{calzetti_dust_2000} spectral windows, modified for the low resolution observations with NIRSpec/PRISM. Furthermore, we also measure the slope as we fit the continuum and emission line model described in \S~\ref{sec:eml_fit} in both fitting the final spectra and bootstrapping the individual 72 exposures (see \S~\ref{sec:eml_fit}). Using all the methods, we measure the  $\beta_{\rm UV}$ = -2.8$\pm$0.1. We note that, similar to the findings of \citep{Baker25_zap}, we do not find a dependence of the $\beta_{\rm UV}$ on the exact spectral windows used to estimate the $\beta_{\rm UV}$. 

We explore the measurement of $\beta_{\rm UV}$ to estimate the escape fraction (f$_{esc}$) based on the correlation described in \citet{Chisholm_escape_frac_2022}:
\begin{equation}
    f_{esc} = (1.3\pm0.6) \times 10^{-4} \times 10^{(1.22\pm0.1)\beta_{UV}}.
\end{equation}
Using this expression, we find the f$_{\rm esc}$ of 0.44$\pm$0.12, consistent with f$_{esc}$ value from the \beagle\ SED fitting (see section below). The incredibly low $\beta_{\rm UV}$ value is steeper than most of the high-z galaxies (see e.g. \citealt{roberts-borsani2024, saxena24}) and more consistent with those found in the napping galaxies such as JADES-GS8-RL-1 \citep{Baker25_zap}.

\subsection{\normalfont\scshape{beagle-agn} \normalfont{Results}}
\label{sec:beagleagnresults}

\begin{figure*}
        \centering
        \includegraphics[width=0.99\textwidth]{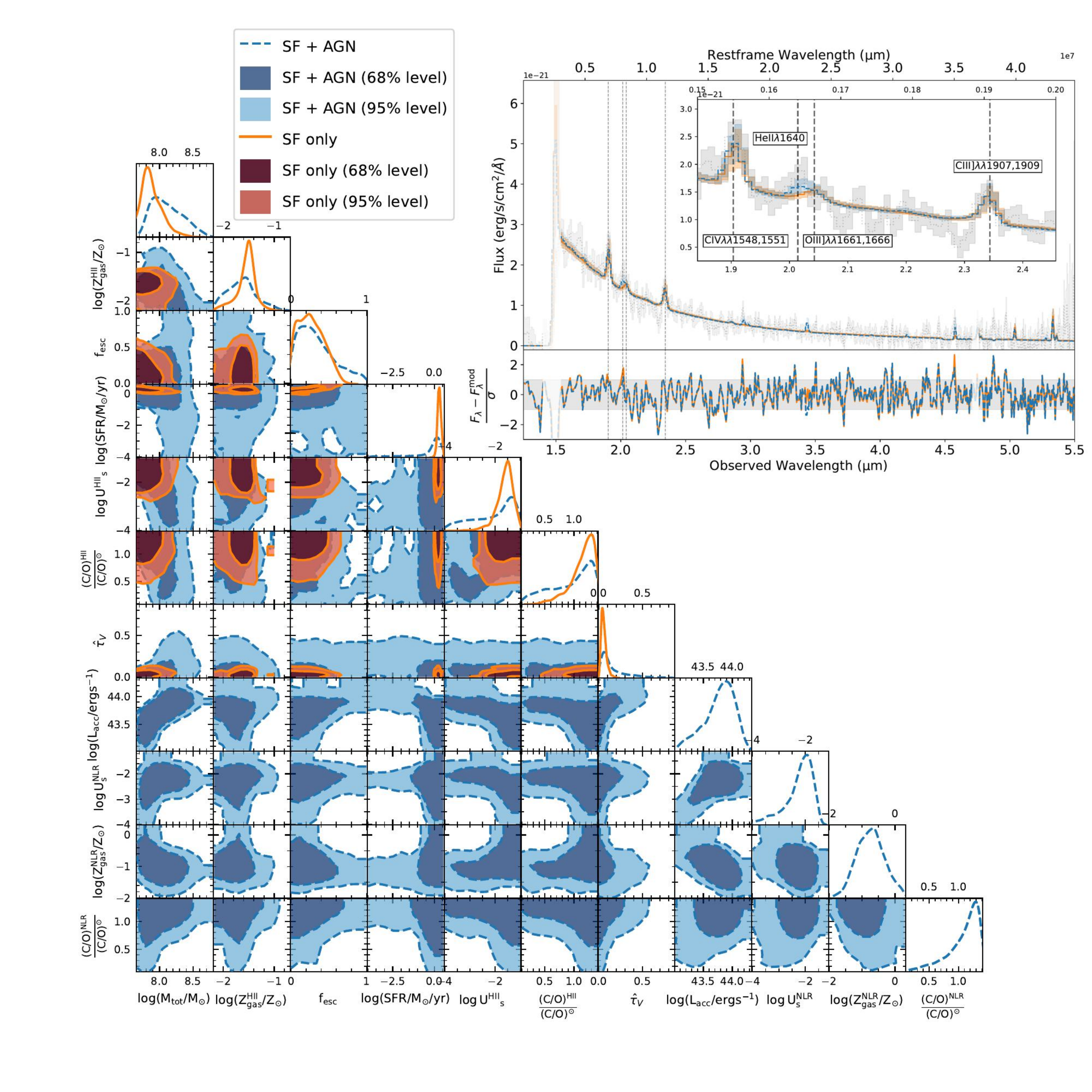}
        
    \caption{\textit{Lower Left}: A triangle plot displaying both \ticks{SF+AGN} (dashed blue line and blue filled contours) and \ticks{SF-only} (solid orange line and orange filled contours) \beagleagn\ fits to \target. Central dark contours in the panels refer to the 1$\sigma$ ($68\%$) credible level, and the outer lighter contours refer to the 2$\sigma$ ($95\%$) credible level of the 2D posterior probability distributions. The uppermost panels for each column display the 1D posterior probability distributions for their corresponding parameters. Parameters shown are: stellar mass $\mathrm{log(M_{tot}/\Msun)}$, \HII\ region metallicity $\mathrm{log(\Zhii/\Zsun)}$, Lyman-continuum escape fraction $\mathrm{f_{esc}}$, star formation rate \sfrconsistent, \HII\ region ionisation parameter \logUsHII, \HII\ region C/O abundance $\mathrm{(C/O)^{HII}}/\mathrm{(C/O)^{\odot}}$, effective $V$-band attenuation optical depth \tauV, accretion disk luminosity $\mathrm{log(\Lacc/ergs^{-1})}$, NLR ionisation parameter \logUsAGN, NLR metallicity $\mathrm{log(\ZAGN/\Zsun)}$ and NLR C/O abundance $\mathrm{(C/O)^{NLR}}/\mathrm{(C/O)^{\odot}}$. \textit{Upper Right}: A plot displaying the marginal SEDs for the same two fits as in the triangle plot, with the \ticks{SF+AGN} fit shown in blue dashed line and the \ticks{SF-only} fit shown in solid orange line; the $95\%$ error region for each of these lines is shown by light blue and light orange shading, respectively. Also included in the upper panel is the measured spectrum in dotted gray line with light gray shading for the $68\%$ error region. Areas that were masked when fitting, namely Ly-$\alpha$ and \NeIII, are indicated by lighter colours with respect to the rest of the spectrum. The $1.5 - 2.0\ \mu\mathrm{m}$ restframe range is shown in a zoom-in insert, which highlights the \CIVall, \HeIIL[1640], \OIIIUvall\ and \CIIIall emission lines. Residuals attributed to these emission lines, as well as the remaining spectral range, are shown on the lower panel. This also contains horizontal grey shading for the $\pm1\sigma$ region and vertical dashed light grey lines assigned to the emission lines highlighted in the zoom-in insert, which extend to the upper panel for visual clarity.}
    \label{fig:beagle_agn_plots}
\end{figure*}

As outlined in Section \ref{sec:beagleSED}, in order to explore additional parameter derivations, we performed spectral fitting of \target\ using \beagleagn\ with priors outlined in Table \ref{tab:fiducialparameters}. 
We show the results of fitting to \target using both the star-forming and NLR models (\ticks{SF+AGN}) and using the star-forming only models (\ticks{SF-only}) in Figure \ref{fig:beagle_agn_plots}.  
In the triangle and marginal plots, contours and lines with blue and orange colours indicate the \ticks{SF+AGN} and \ticks{SF-only} fit, respectively. Panels in the triangle plot of Fig. \ref{fig:beagle_agn_plots} which display 2D probability distributions for SF-only related parameters, naturally include the \ticks{SF+AGN} and \ticks{SF-only} fits, as these both share those parameters in common. Parameter combinations which include pairings of NLR and \HII\ region parameters result in panels including just the fiducial \ticks{SF+AGN} in comparison.\\

At first glance, the posterior probability distributions for the \ticks{SF+AGN} fit are typically less constrained in comparison to those of the \ticks{SF-only} fit, although with posterior peaks often overlapping with those of the \ticks{SF-only} fit. This is to be expected given the increased number of parameters in the SF+AGN fit (and therefore increased number of degeneracies), as well as the spectrum of \target\ not including many extremely high ionisation potential emission lines. The spectrum indeed includes emission lines such as \CIVall\ and \CIIIall, however, in the absence of significant emission lines with even higher ionising potential that cannot be explained by star-formation alone (for example \NeV and \NeIV or significant \HeIIL[1640] emission), 
the NLR parameters become more challenging to constrain.

Comparing the SF+AGN and the SF-only fit, all parameters in common are consistent to within $\pm1\sigma$, except for $\log(\MtotInLog/\Msun)$. These consistencies largely owe to the lesser constrained posteriors for the SF+AGN fit as discussed previously. The parameter with the largest statistical difference between the two fits is indeed $\log(\MtotInLog/\Msun)$, for which the \ticks{SF-only} fit is $\sim1.9\sigma$ lower than that of the \ticks{SF+AGN} fit. This occurs in concert with the SFR increasing in the \ticks{SF-only} fit (\sfrconsistent$_{\mathrm{SF-only}} = 0.32^{+0.11}_{-0.10}$) with respect to the \ticks{SF+AGN} fit (\sfrconsistent$_{\mathrm{SF+AGN}} = -0.09^{+0.36}_{-0.13}$) in order to reproduce the emission line fluxes, thereby allocating more luminosity to the younger stellar populations and in turn lowering the stellar mass. When considering the maximum stellar ages in both fits, $\mathrm{log(t_{max}/yr)_{SF-only} = 7.37^{+0.27}_{-0.26}}$ and $\mathrm{log(t_{max}/yr)_{SF+AGN} = 7.50^{+0.32}_{-0.32}}$, both suggest that the stellar contribution to the SED of \target\ is dominated by younger stars, again indicating a younger stellar population. 

As described above, the \target\ emission line ratios do not rule out the AGN nature of this target. Given that this source may host an AGN, this work's SF+AGN fit derives some parameter estimates relating to the proposed AGN component. This includes a sub-solar NLR metallicity ($\mathrm{12 + log(O/H)}^{\mathrm{NLR}} = 7.9^{+0.4}_{-0.4}$), a moderate NLR ionisation parameter ($\logUsAGN = -2.1^{+0.4}_{-0.4}$) and a moderate accretion disk luminosity ($\mathrm{log(\Lacc/ergs^{-1})} = 43.8^{+0.2}_{-0.2}$). The latter places \target\ within the range of bolometric luminosity estimates from \citet[41.5 $\lesssim$ log($\mathrm{L_{bol} / ergs^{-1}}$) $\lesssim$ 44.5]{Scholtz23AGN} and below the derived accretion disk luminosity of the Type-II AGN candidate from \citet[log($\mathrm{L_{bol} / ergs^{-1}}$) = $45.19^{+0.12}_{-0.11}$]{Silcock24}. 

With our \beagle\ and \beagleagn\ fitting, we were able to explore model grids covering a range of C/O abundances ($\mathrm{(C/O)}/\mathrm{(C/O)^{\odot}} \in [0.1, 1.4]$) for both the \HII\ and potential NLR components. When fitting with a SF component only, a super-solar C/O abundance is required to reproduce the observed carbon emission lines in the UV ($\mathrm{(C/O)^{HII}_{SF-only}}/\mathrm{(C/O)^{\odot}} = 1.20^{+0.15}_{-0.15}$). Even when including an AGN component, the derived NLR C/O abundance is also super-solar, though consistent with solar to 1$\sigma$ ($\mathrm{(C/O)^{NLR}}/\mathrm{(C/O)^{\odot}} = 1.12^{+0.22}_{-0.23}$), with the NLR component contributing $65^{+32}_{-38}\%$ and $67^{+32}_{-50}\%$ to the \CIIIall\ and \CIVall\ emission lines, respectively. The \HII\ region C/O abundance is less constrained in this fit ($\mathrm{(C/O)^{HII}_{SF+AGN}}/\mathrm{(C/O)^{\odot}} = 0.95^{+0.36}_{-0.13}$). Therefore, regardless of if \target\ is considered to be a definite Type-II AGN, or instead a SF dominant object, carbon enhancement is the best explanation of the strength of the observed UV lines.

The upper right plot of Fig. \ref{fig:beagle_agn_plots} includes a panel displaying the residuals from the SF+AGN and SF-only fits to \target. With the horizontal grey shading denoting the $\pm1\sigma$ region, the modelled spectra of both SF+AGN and SF-only fits appear statistically sound overall, especially when noting that the unmasked residuals rarely go beyond the $\sim2\sigma$ level. Comparing the \multinest\ output statistics between the fits, we note the Log-Evidence values were $-0.877 \pm\ 0.339$ and $-0.875 \pm\ 0.282$ for the SF-only and SF+AGN fits, respectively, affirming their overall similarity in performance. Focusing on the $1.5 - 2.0\ \mu\mathrm{m}$ rest-frame wavelength range, however, we can consider some emission lines of interest in more detail. Here we note the \CIIIall\ and \CIVall\ lines are well modelled by both SF+AGN and SF-only fits. The highest ionisation potential of the carbon emission lines present (47.89 eV, \CIVall) can be difficult for standard stellar populations to reproduce \citep{Feltre16}; however we suspect its reproduction by the SF-only fit to be facilitated by a compound effect of the super-solar C/O abundance, and a hard ionising field ($\mathrm{(\logUsHII)_{SF-only}} = -1.55^{+0.25}_{-0.26}$) from the stellar population (which was modelled to extend up to $\mathrm{300 M_{\odot}}$). In the same zoom-in insert is \HeIIL[1640], which is a tentative feature in the spectrum at $\lesssim$ 2.7$\sigma$. Reaching this ionising potential ($\sim54.42$ eV) is more difficult for SF-only modelling, hence any future SF-only fit of deeper data would not be able to reproduce a confirmed \HeIIL[1640] emission. Therefore a deeper PRISM or R1000 spectra is essential to confirm the nature of \target. We caution, however, that our stellar models \citep{gutkin_modelling_2016} at present do not include X-ray binary prescriptions, which may otherwise have improved the modelling of \HeIIL[1640] in the absence of an AGN component. 


\subsection{ISM properties of \target}\label{sec:ISM_prop}

The detection of multiple emission lines allows us to study the ISM properties. In this section, we are assuming that the primary source of ionisation in this galaxy is due to star formation rather than AGN. To perform a more detailed modelling of the ISM conditions, ideally, we need high-resolution observations to resolve the \CIIIall, \OIIall doublets to resolve the doublets sensitive to density and temperature of the ISM (e.g., \citealt{berg_chemical_2019, kewley_understanding_2019}). However, most of our detections are in the PRISM observations with low resolution (R$\sim$30-300), which leaves these emission lines spectrally unresolved. Therefore, we are forced to use scaling relationships and calibrations used in the literature to derive ISM properties such as metallicity and ionisation parameter.

We take advantage of the detection of the H$\gamma$ emission line to estimate the SFR of \target independently of the SED fitting. In our estimate, we assume Case-B ratio for the conversion of H$\gamma$ to H$\alpha$  of 6.085 (H$\gamma$/H$\beta$ = 0.470 and H$\alpha$/H$\beta$=2.86). Given the steep UV-slope and the SED fitting showing no evidence for dust obscuration, we do not correct our Balmer lines for dust obscuration. We use the conversion from \citet{Kennicutt12} to convert H$\alpha$ (H$\gamma$) luminosity to SFR. We finally estimate the log(SFR/\Msun yr$^{-1}$) from H$\gamma$ 0.72$\pm$0.15. However, using calibrations for low metallicity galaxies from \citet{Reddy_2018} and \citet{Theios19}, we estimate the log(SFR/\Msun yr$^{-1}$) of 0.50$\pm$0.15 and 0.35$\pm$0.17, respectively. Given the range of estimated SFR, we chose the central value using calibration from \citet{Reddy_2018}. However, we note that there is systematic uncertainty on the SFR of $\sim$0.25 dex from the different calibrations alone.

We can estimate the ionisation parameter using the detected carbon lines - \CIIIall and \CIVall, following the procedure outlined by \citep{Mingozzi23}, who calibrated this ratio via the ionisation parameter estimated through the \OIIIall/\OIIall measurements. Similarly to the \OIIIall/\OIIall, the \CIVall/\CIIIall is not sensitive to the abundance of the individual abundances of different elements. The estimated ionisation parameter from the \CIVall/\CIIIall is log U = -1.9$\pm$0.1, with a systematic uncertainty of 0.3. Furthermore, we used calibrations from \citet{witstok_lensed_z5_2021} to estimate the log U from the \NeIII/\OIIall ratio of -1.93$\pm$0.12, consistent with that estimated from \CIVall/\CIIIall.

\begin{figure}
        \centering
	\includegraphics[width=0.99\columnwidth]{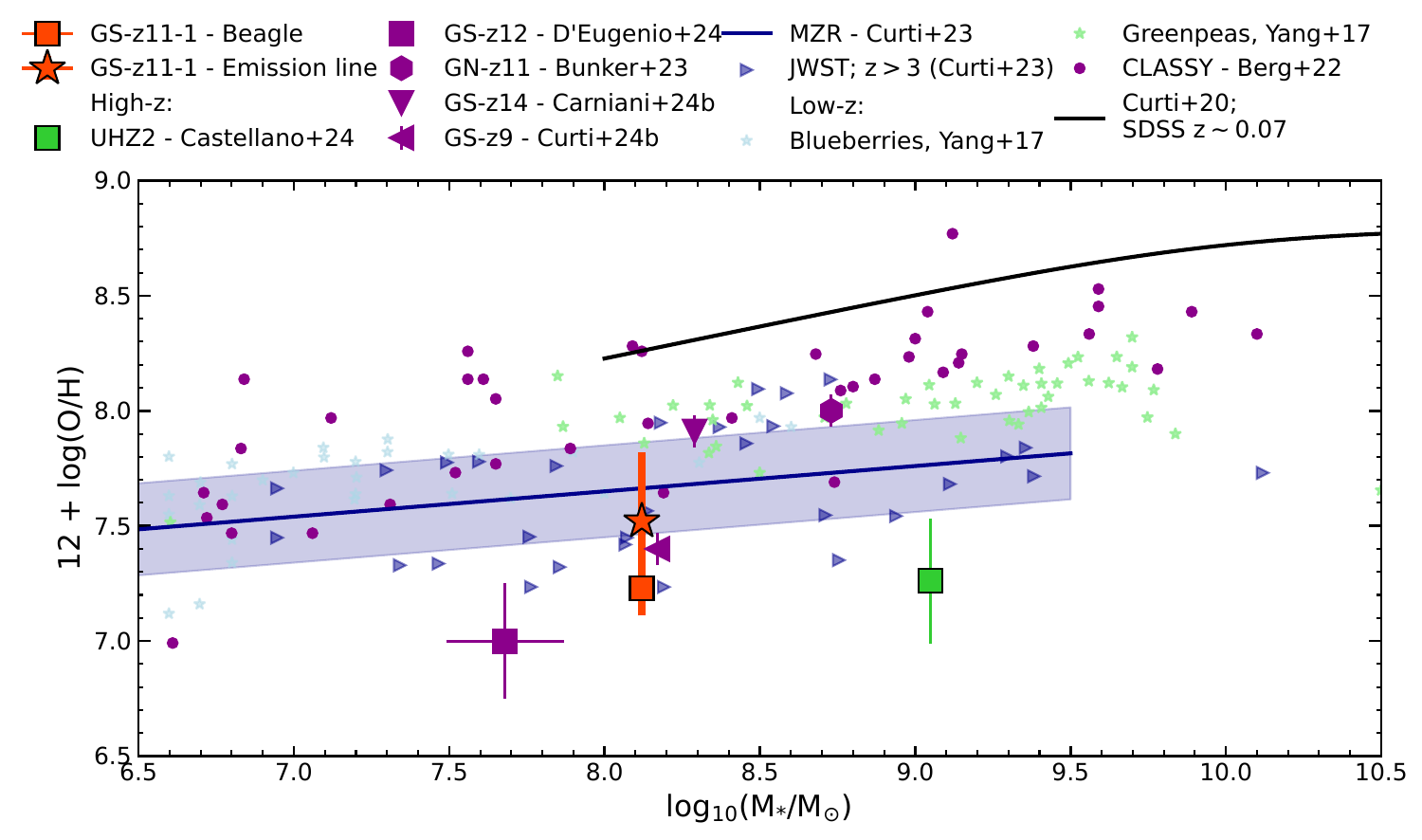}
    \caption{Mass-metallicity relation (MZR) plot of our target with other sources at z$>$9 from the literature \citep[various symbols][]{Bunker23gnz11, DEugenio24z12, Castellano24, Carniani2025} and MZR at z$>$3 from \citet{curti_jades_mzr_2024} (blue shaded region and blue circles), SDSS \citep[black line][]{curti_massmetallicity_2020} and local high-z analogues \citep[blueberries and greenpeas as blue green stars][]{yang_blueberries_2017} and from the CLASSY survey \citep{Berg22}.}
    \label{fig:metal}
\end{figure}

In order to estimate the metallicity, we use the Ne3O2 = \NeIIIL[3869]/\OIIall ratio, which is considered a possible tracer of oxygen abundance in the absence of other optical lines \citep[e.g.][]{maiolino_amaze_2008}. We chose calibrations from \citet{Curti20}, which extends to lower metallicity expected in our object. We estimate a metallicity of 12 + log (O/H) = 7.5$_{-0.4}^{+0.4}$, corresponding to Z= 0.1$^{+0.1}_{-0.09}$ Z$_{\odot}$. As mentioned by \citet{Napolitano2024}, the \OIIall may be suppressed by collisional de-excitation at high density ($\sim 8,000$ cm$^{-3}$), and hence we need to explore additional calibrations. We utilize the detection of the \CIIIall emission line and calibrations from \citet{Mingozzi23}. Using the EW of the \CIIIall we estimated 12+log(O/H) of 8.0$\pm$0.3, consistent with the Ne3O2 metallicity estimated above. Overall we can conclude that the metallicity of our object is 5-20\% Z$_{\odot}$.

In Figure~\ref{fig:metal}, we compare the metallicity of our target with the rest of the high redshift sources in the literature \citep{Bunker23gnz11, DEugenio24z12, Castellano24, Curti24_9.4, Carniani2025}, along with local high redshift analogues such as green peas and blueberries. The \target is consistent with the mass metallicity plane derived from \citet{curti_jades_mzr_2024} for high-z galaxies, given its stellar mass derived from the \beagle\ SED fitting. This shows that this is a low metallicity galaxy with high ionisation parameter, typical of strong \CIVall emitters such as GHZ2, RXCJ2248-ID and GN-z11 \citep{Castellano24, topping_z6_lens_2024, Maiolino23gnz11}.

\begin{figure}
        \centering
        \includegraphics[width=0.95\columnwidth]{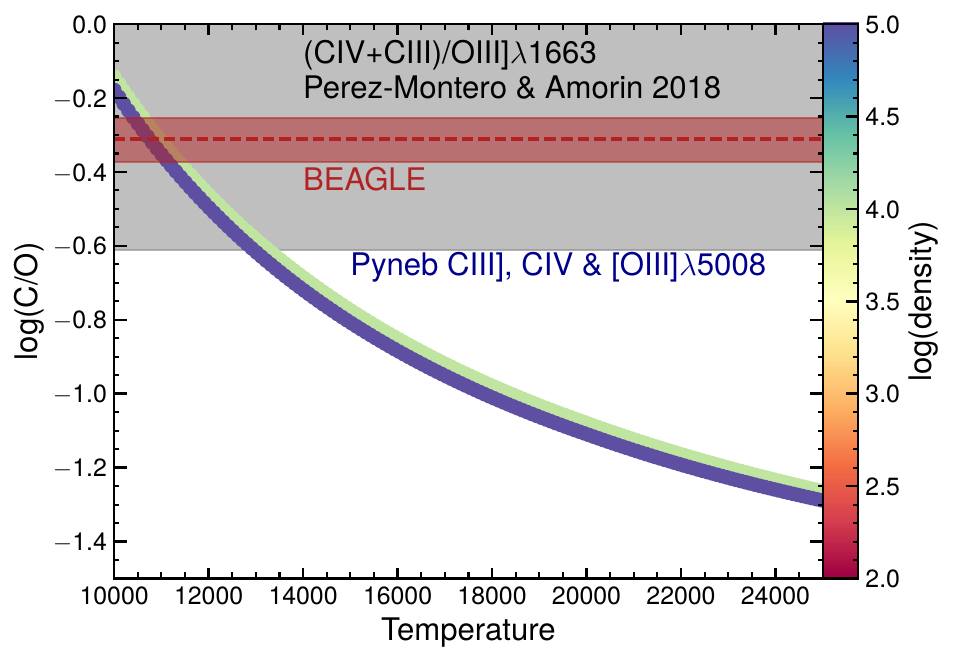}
	\includegraphics[width=0.95\columnwidth]{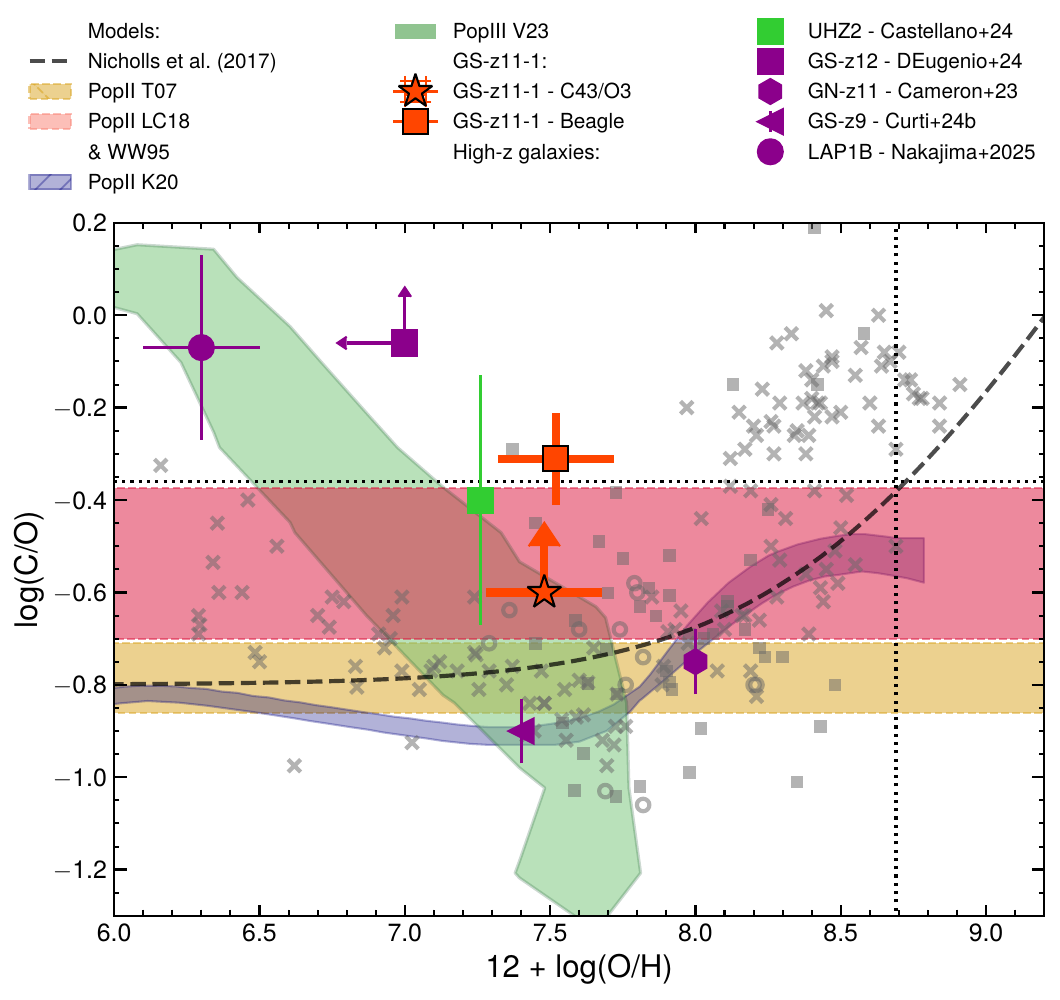}
    \caption{Top panel: Our C/O estimate using the two separate methods. The grey region shows the lower limit by calibration from Perez-Montero \& Amorin using \CIIIall, \CIVall and upper limit on \OIIIL[1663]. The coloured lines show the C/O estimate from \texttt{Pyneb} using the \CIIIall and \OIIIL[5008] estimate (the densities of 10$^{2-3}$ cm$^{-3}$ are obscured by the 10$^{4}$ cm$^{-3}$ line). The two methods agree for the electron temperature of $\lesssim$14 000 K. Bottom panel: C/O vs O/H of our target (red stars) and comparison to values in the literature from z$>9$ galaxies and local metal poor galaxies. We show theoretical yields of Pop II \& III supernovae (black dashed line; yellow, red, blue and green shaded regions; \citealt{woosley+weaver1995, tominaga_2007, heger+woosley2010,limongi+chieffi2018, kobayashi_origin_2020, vanni+2023}). The dotted line shows the solar abundances as defined by \citet{gutkin_modelling_2016}.
    }
    \label{fig:CO_plot}
\end{figure}

\subsection{C/O and N/O abundances}\label{sec:CO_abud}

With \target in this work, there are now three sources with detection of carbon emission lines at z$>$11. Given the detection of both oxygen and carbon lines, we can constrain the C/O abundance ratio and investigate the enrichment paths of this galaxy.

To estimate the C/O, we use two separate approaches, using different emission lines. Firstly, we use the calibration described in \citet{perez-montero_using_2017}, specifically their eq. 3, using the detection of \CIVall, \CIIIall and upper limit on \OIIIL[1663]. Using this calibration, we estimate the lower limit on log(C/O) of $>$-0.6. The second method was adapted from \citet{Castellano24} using \texttt{PyNeb} \citep{luridiana_pyneb_2012, luridiana_pyneb_2015}, specifically using the function \texttt{getIonAbundance}. We consider a range of electron densities - [10$^{2}$,10$^{3}$,10$^{4}$,10$^{5}$] cm$^{-3}$ and electron temperature of 1-3$\times 10^{4}$ K. We assume that the C/O abundance can be estimated from the C$^{2+}$ and O$^{2+}$ abundance using the \CIIIall and \OIII. As we do not have a detection of \OIIIUvall, we take advantage of relation between \NeIIIL[3869], \OIIall and \OIIIL[5008] from \citet{witstok_lensed_z5_2021}:
\begin{equation}
    \log_{10} \left( \frac{\NeIII[3869]}{\OIIall} \right) = 0.9051 \log_{10} \left( \frac{\OIIIL[5008]}{\OIIall} \right) - 1.078,
\end{equation}
to estimate the flux of \OIIIL[5008].
The ionisation potential of O$^{2+}$ is 7eV higher than C$^{2+}$, and as such, we need to apply the ionization correction factor (ICF). We use the calibration from \citet{berg_chemical_2019}, which depends on the ionisation parameter and metallicity. For the metallicity of this object and ionisation parameter, we estimate the ICF of 1.39. We show the C/O vs temperature and density in the top panel of Figure \ref{fig:CO_plot}. The estimate of the C/O based on this method ranges from -0.18 to -1.22, an order of magnitude difference. Overall, the C/O estimates are in agreement for the electron temperature of $<$14,000 K, which is in agreement with the very loose constraint from the \OIIIL[4363] upper limit of $<$27,000 K. 

In the bottom panel of Figure \ref{fig:CO_plot} we compared the C/O abundance of \target with other sources at high redshift from the literature \citep{Cameron23gnz11, Castellano24, Curti24_9.4, Nakajima2025} as well as various models theoretical yield models (\citealt{woosley+weaver1995, tominaga_2007, heger+woosley2010,limongi+chieffi2018, kobayashi_origin_2020, vanni+2023}). Our galaxy has one of the highest C/O enrichment in the literature at high redshift, consistent with enrichment by either PopIII or PopII stars. We note that the lower C/O and higher O/H compared to GS-z12 \citep{DEugenio24z12} could potentially be due to dilution of the PopIII encirhcment by subsequent generations of stars after the initial burst PopIII stars. Hence, we are most likely looking at a galaxy at a later evolutionary stage compared to GS-z12.

\begin{figure}
        \centering
	\includegraphics[width=0.95\columnwidth]{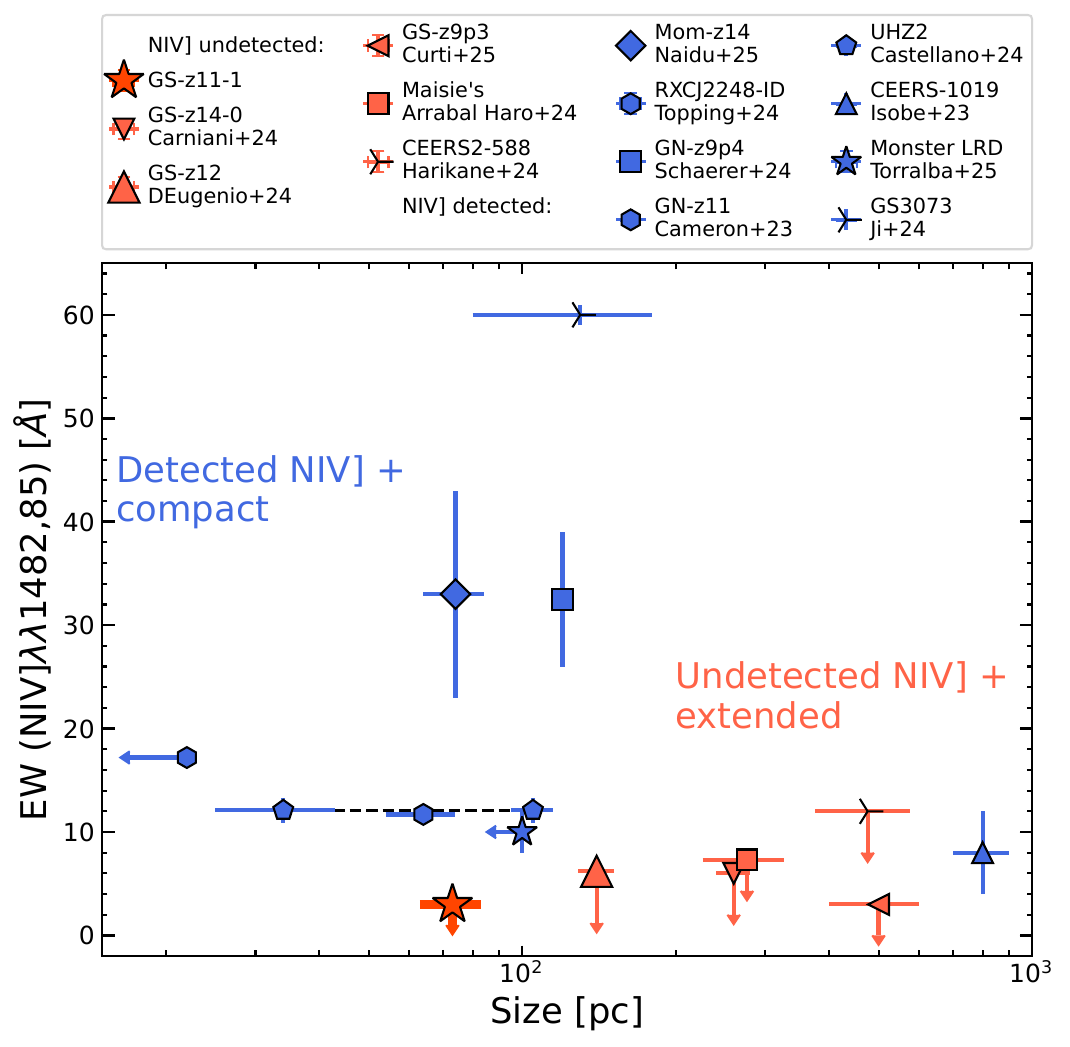}
    \caption{\NIVall equivalent width vs Size of \jwst observed galaxies at z$>$6. We split the sample from the literature those with (blue) and without (orange) detected \NIV emission line: Mom-z14 \citep{Naidu2025}, RXCJ2248-ID \citep{topping_z6_lens_2024}, GS-z14 \citep{carniani_z14_2024}, GN-z9p4 \citep{Schaerer24}, GS-z12 \citep{DEugenio24z12}, GN-z11 \citep{Cameron23gnz11}, UHZ2 \citep{Castellano24}, GS-z9p3 \citep{Curti2025}, CEERS-1019 \citep{Isobe23}, UNCOVER-45924 \citep{Torralba25}, CEERS2-588 \citep{Harikane23} and GS3073 \citep{ji_nitrogen_AGN_z5_2024}. We colour the comparison sample according \NIV detection, similar to \citet{Naidu2025}.
    }
    \label{fig:Nitrogen_plot}
\end{figure}

Despite the strong detection of carbon, neon and oxygen lines, we do not see any evidence of a detection of nitrogen lines such as \NIII and \NIV. Using the procedure outlined above for the C/O using \texttt{pyneb}, we estimated the upper limit on the log(N/O) abundance of $<-0.3$. In Figure \ref{fig:Nitrogen_plot}, we compare the \NIVall equivalent width and size of our galaxy and a sample of galaxies from the literature. We find that the equivalent widths of the \NIII and \NIV lines of \target ($<2.3$ and $<3.6 \AA$, respectively) are at least a factor of three lower than the nitrogen loud (or nitrogen strong) galaxies in the literature \footnote{Exact definition of nitrogen loud/strong galaxies in the literature is uncertain, and most likely corresponds to whether \NIV or \NIII is detected in the spectra. We adopt this approach in this work.}, despite its size being consistent with that of Mom-z14 \citep{Naidu2025} and GN-z9p4 \citep{Schaerer24}. 

As shown recently by \citet{Ji_2025_GC, Senchyna_gnz11_2024}, the nitrogen enhancement in the above compact galaxies could potentially trace an early evolutionary stage of galaxies where star formation is dominated by dense star clusters. These star clusters are chemically self-enriched and are progenitors of globular clusters observed in the local Universe (although in these galaxy cores are likely to merge to form nuclear star clusters). In these systems, the stars born out of the gas enriched by the first-generation stars exhibit significantly enhanced nitrogen abundances. These later generation stars, however, typically do not exhibit enhanced C/O as carbon is turned into nitrogen through the CNO cycle. Instead, the high C/N ($>0.3$ dex) and low O/H of \target is more similar to that of carbon-enhanced metal poor stars \citep[CEMPs,][]{Beers_2005} and might trace an even earlier chemical evolution stage, although their origins remain unclear \citep{Norris_2013}. At the time of the writing there are three sources with enhanced C/O at z$>10$ (\target,  GS-z12, LAP1B; \citealt{DEugenio24z12, Nakajima2025}), all of them without any direct evidence for enhanced N/O, despite their compact sizes.  With \target and GS-z12, it is possible we are looking at the evolutionary stage of high-z galaxies with intense nuclear star-formation preceding the stage of nitrogen enhanced galaxies (i.e., before the star clusters became significantly self-enriched), potentially tracing the birth environments of CEMPs.

\subsection{Dynamical masses}\label{sec:Mdyn}

As noted above, we detect \CIIIall emission line at 3.2$\sigma$ significance in the R1000 spectrum. Fitting a single Gaussian profile to the \CIIIall doublet showed an intrinsic FWHM of the emission line of 520$^{+193}_{-170}$ \kms, indicating that the line is resolved in the grating observations with LSF FWHM of 170 \kms. Therefore, we fit the \CIIIall emission line with two Gaussian components, which corresponds to each emission line in the doublet, each having the same FWHM and redshift. We obtain the FWHM of 376$^{+242}_{-164}$ \kms. The ratio of [CIII]$\lambda$1907/CIII]$\lambda$1909 emission lines is unconstrained in the fit, therefore, we are unable to constrain the electron density based on the \CIIIall doublet. 

Given that we spectroscopically resolve the emission line in the R1000 observations, we are now able to loosely constrain the dynamical mass (M$_{\rm dyn}$), assuming that the system is close to virialization, and can compare it to the stellar mass from the \beagle\ SED fitting. 
We estimated the dynamical mass following the approach described by \citet{Ubler_AGN_z5_2023} using the equation:
\begin{equation}
    M_{\rm dyn} = K(n) K(q) \frac{\sigma^{2}R_{e}}{G}
\end{equation}
where K(n) =  8.87 -- 0.831n + 0.0241n$^2$ and with S\'{e}rsic index n, following \citet{Cappellari2006}, K(q) = [0.87 +
0.38e$^{-3.71(1-q)}]$$^{2}$, with axis ratio q following \citet{van_der_Wel2022}, $\sigma$ is the integrated stellar velocity dispersion, and R$_{e}$ is the effective radius.

It is important to note that the above calibration is based on the stellar kinematics of massive galaxies at lower redshift ($z=0.8$, $\Mstar= 10^9\text{--}10^{11}~\msun$). However, as noted in \citet{Carniani2025}, a similar calibrations such as \citep[e.g.,][]{Wisnioski:2018,Cappellari:2013} provides similar answers, within 0.3~dex \citep[see e.g.][]{Marconcini+2024}. We apply a correction of 
-0.18~dex following \citet{Ubler_AGN_z5_2023}, since the stellar dispersion tends to be lower than the ionised gas velocity dispersion \citep{Bezanson:2018}. We adopt a size (R$_{e}$) of 70$\pm$10 pc from the Forcepho fitting. For the serseic index (n), we adopt a range of values for the S\'{e}rsic index and axis ratio (n=0.8--2 and q=0.3--1).
Thus, we estimate a log (M$_{\rm dyn}/\msun$) = 9.0$\pm0.5$.

This estimated value of M$_{\rm dyn}$ is $\sim$1.2dex higher than the stellar mass estimated from \beagle\ SED fitting, implying a fraction of gas mass and dark matter in this galaxy of 65--95\%. This would imply a large gas reservoir in this galaxy, opposite to the recent gas estimates in another galaxy, GS-z14 \citep[][]{Carniani2025, Scholtz+2025}. However, we note that this measurement is based on a single, barely spectrally resolved emission line doublet, and future high-spectral-resolution \jwst or ALMA observations are necessary to further constrain the kinematics and dynamical mass of this system. 

\section{Summary and discussion}\label{sec:summary}

We report the spectroscopic confirmation of \target at z=11.18, near Cosmic Dawn, based on new spectroscopic observations with NIRSpec/MSA in the JADES programme. Using new data reduction, we were able to expand the nominal wavelength coverage of NIRSpec/PRISM observations to a maximum wavelength to 5.5 $\mu$m. We report detection of multiple emission lines from the rest frame UV to the optical: \CIVall, \CIIIall, \OIIall, \NeIIIL[3869] and H$\gamma$. This target, along with a growing number of detections of emission lines at z$>10$ \citep{Bunker23gnz11, DEugenio24z12, carniani_z14_2024} shows the ability to detect emission lines even in high-z galaxies with \jwst/NIRSpec-MSA. 

The detection of several emission lines has enabled us to perform a detailed comparison with photo-ionisation models \citep{Feltre16, gutkin_modelling_2016, Nakajima22} to search for the dominant source of ionisation in this galaxy.
On multiple UV line diagnostics
\target can be reproduced by both star-formation and AGN activity, outlining the difficulty of identifying AGN at high redshift \citep{Scholtz23AGN, Castellano24, Maiolino23gnz11}. However, we note that if future observations confirm the 2.5$\sigma$ tentative detection of the \HeIIL[1640] line, then this could confirm \target as an AGN host galaxy. 

We modelled the galaxy's SED with \beagle\ SED fitting code. Given the ambiguity in the source of ionisation we modelled it both as a star-forming galaxy and as an AGN host galaxy with a modified version of \beagle\ called \beagleagn. The galaxy is equally well fitted with both AGN and star-forming models. We have shown that the SF model is a better fit model, due to fewer free parameters.


We are able to constrain the stellar mass to log($M_{\rm tot}$/M$_{\odot}$) = 7.84$^{+0.13}_{-0.11}$. Furthermore, we estimated a SFR of 2.1$\pm$0.5 \Msun yr$^{-1}$, which is well in agreement with the estimated value from H$\gamma$ of 4.1$\pm$1.2 \Msun yr$^{-1}$, under an assumption that all of the flux in the H$\gamma$ line is due to ionisation from star-formation and A$_{\rm V}$=0. Furthermore, both the steep UV slope and the \beagle\ SED fitting indicates a large escape fraction in this galaxy in the range of 30--45 \%. 

We estimated the physical ISM properties using the suite of emission lines detected or constrained by the PRISM spectrum. Using standard calibration from UV and optical emission lines from \citet{Mingozzi23} and \citet{witstok_lensed_z5_2021} we estimated the ionisation parameters (log U) of -1.9$\pm$0.1 and -1.93$\pm$0.12, respectively, within an excellent agreement. Using metallicity calibrations from \citet{Curti20} and \citet{Mingozzi23}, we estimated the metallicity (12+ log(O/H)) of 7.5-8.0 (5-20\% Z$_{\odot}$), indicating a metal-poor high ionisation ISM conditions, similar to other high-z galaxies \citep{Castellano24, Curti24_9.4, Bunker23gnz11}. We compared \target on the mass metallicity plane (see Figure \ref{fig:metal}) to other z$>$9 galaxies, showing that the metallicity of \target is consistent with other galaxies at z$>$10. 

The detection of the strong carbon emission lines compared to the non-detection of \OIIIUvall and weak \OIIall indicates high C/O abundance in this system. We investigate the abundance using multiple different approaches to estimate the C/O.
in Fig. \ref{fig:CO_plot}. Using the calibration from \citet{Perez-montero17} based on the carbon lines along with an upper limit on \OIIIUvall gives a lower limit on the log(C/O) of -0.6, consistent with the estimate from \beagle\ SED fitting of -0.28$\pm$0.15. Although this is a higher C/O abundance ratio relative to other sources at high-z  \citep{arellano-corodva_2023, jones_CO_z6_2023, stiavelli+2023,Castellano24, Cameron23gnz11, Curti24_9.4}, we find that the high C/O in this galaxy is still consistent with PopII core-collapse SNe yields from \citet{woosley+weaver1995, lian_mass-metallicity_2018}; however, given the high C/O we cannot exclude some contribution to the carbon enrichment from PopIII stars. Despite the compactness (73$\pm$10 pc) and high C/O abundances, we see no evidence of an increased equivalent widths of UV nitrogen lines such as \NIII and \NIV or an increase in N/O abundance. Given the evidence that the increase of N/O is linked to the second generation of stars in a globular cluster and the lower C/O compared to other extreme C/O sources \citep{DEugenio24z12, Nakajima2025}, we hypothesise that the previous carbon enrichment from PopIII stars has been diluted by the later population of stars. However, it is too early in the evolution of the galaxy for the second generation of globular cluster stars to make an appearance and enrich this galaxy with nitrogen, possibly witnessing the birth of CEMP stars in the early Universe. 

\section*{Acknowledgements}
This work is based on observations made with the NASA/ESA/CSA James Webb Space Telescope. The data were obtained from the Mikulski Archive for Space Telescopes at the Space Telescope Science Institute, which is operated by the Association of Universities for Research in Astronomy, Inc., under NASA contract NAS 5-03127 for JWST. These observations are associated with program 1287.
We are grateful to S. Salvadori and I. Vanni for providing the chemical enrichment tracks for some Population III scenarios.
JS, RM, FDE, XJ and GCJ acknowledge support by the Science and Technology Facilities Council (STFC), ERC Advanced Grant 695671 ``QUENCH" and the UKRI Frontier Research grant RISEandFALL. RM also acknowledges funding from a research professorship from the Royal Society.
MSS acknowledges support by the Science and Technology Facilities Council (STFC) grant ST/V506709/1.
ECL acknowledges support of an STFC Webb Fellowship (ST/W001438/1).
The Cosmic Dawn Center (DAWN) is funded by the Danish National Research Foundation under grant DNRF140.
S.C acknowledges support from the European Union (ERC, WINGS,101040227)
WMB gratefully acknowledges support from DARK via the DARK fellowship. This work was supported by a research grant (VIL54489) from VILLUM FONDEN.
SA acknowledges grant PID2021-127718NB-I00 funded by the Spanish Ministry of Science and Innovation/State Agency of Research (MICIN/AEI/ 10.13039/501100011033)
AJB and JC acknowledges funding from the "FirstGalaxies" Advanced Grant from the European Research Council (ERC) under the European Union’s Horizon 2020 research and innovation programme (Grant agreement No. 789056).
DJE is supported as a Simons Investigator and by JWST/NIRCam contract to the University of Arizona, NAS5-02015.  Support for program \#3215 was provided by NASA through a grant from the Space Telescope Science Institute, which is operated by the Association of Universities for Research in Astronomy, Inc., under NASA contract NAS 5-03127.
YI is supported by JSPS KAKENHI Grant No. 24KJ0202.
PGP-G acknowledges support from Spanish Ministerio  de  Ciencia e Innovaci\'on MCIN/AEI/10.13039/501100011033 through grant PGC2018-093499-B-I00.
BER acknowledges support from the NIRCam Science Team contract to the University of Arizona, NAS5-02015, and JWST Program 3215.
H\"U acknowledges funding by the European Union (ERC APEX, 101164796). Views and opinions expressed are however those of the authors only and do not necessarily reflect those of the European Union or the European Research Council Executive Ag
The research of CCW is supported by NOIRLab, which is managed by the Association of Universities for Research in Astronomy (AURA) under a cooperative agreement with the National Science Foundation.
CNAW acknowledges JWST/NIRCam contract to the University of Arizona NAS5-02015
JW gratefully acknowledges support from the Cosmic Dawn Center through the DAWN Fellowship. The Cosmic Dawn Center (DAWN) is funded by the Danish National Research Foundation under grant No. 140.
The authors acknowledge use of the lux supercomputer at UC Santa Cruz, funded by NSF MRI grant AST 1828315.

\section*{Data Availability}

The datasets were derived from sources in the public domain: JWST/NIRSpec MSA and JWST/NIRCam data from MAST portal - \url{https://mast.stsci.edu/portal/Mashup/Clients/Mast/Portal.html}.



\bibliographystyle{mnras}
\bibliography{mybib,mirko_abundances} 



\appendix

 \section*{Appendix A: Comparison of observations and NIRCam}\label{sec:obs_scaling}

\begin{figure*}
        \centering
	\includegraphics[width=0.8\textwidth]{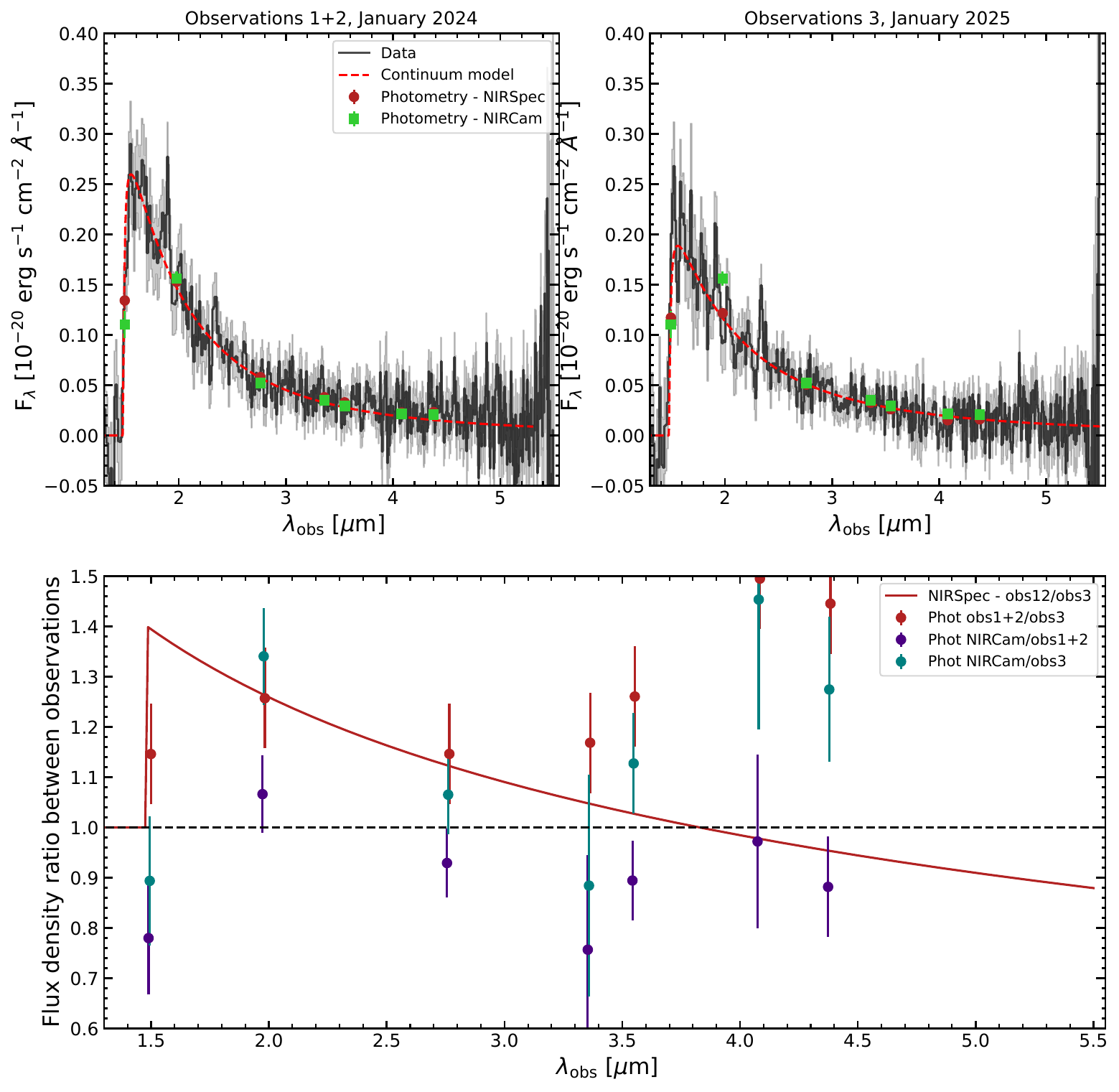}
    \caption{ Comparison of the NIRSpec-MSA PRISM calibration between individual visits and NIRCam imaging. Top panels: Comparison of visits 1+2 from January 2024 (left) and visit 3 from January 2025 with different slit placement (right). While the Visits 1+2 agree well with NIRCam photometry, there is a significant offset in F200W filter between the visit 3 and NIRCam photometry. Bottom panel: Compairson of the mock photometry from NIRSpec PRISM observations and NIRCam for different visits. The red solid line indicates the ratio between the best-fit models of visits 1+2 and visit 3. We use the ratios to normalize the visit 3 to be consistent with visits 1+2.}
    \label{fig:exp}
\end{figure*}

In this section of the appendix, we compare the flux calibrations between the different multiple NIRSpec-MSA visits and NIRCam imaging. We combine the spectra from visits 1 and 2 taken using the same MSA configuration on January 2024 (see slit positions on Figure \ref{fig:Spectrum}) and visit 3 with a separate MSA configuration on January 2025 and we show the combined spectra in the top row of Figure \ref{fig:exp}. 

The January 2025 (visit 3) has a significant discrepancy at the 2$\mu$m region between the PRISM spectrum and the NIRCam photometry. For more quantitative evaluation we extracted the synthetic photometry from the PRISM spectrum. While all other filters are consistent within 1$\sigma$, we see that the F200W filter flux in the PRISM is factor of $\sim$1.5 lower than in the NIRCam photometry and the visits 1 and 2. We plot the ratios of the of the synthetic photometry from PRISM between the observations in the bottom panel of Figure \ref{fig:exp}. Furthermore, we fitted a powerlaw model with a IGM absorption (fully described in \S~\ref{sec:eml_fit}) and we plot the ratio of the continuum best fit of the Visit 1+2 and Visit 3. We see that the deficiency of flux at 2$\mu$m causes the the UV slope to change from the -2.9$\pm$0.1 (Visit 1+2) to -2.5$\pm$0.1 in Visit 3. The fact that the Visit 1+2 agrees with NIRCam photometry establishes that those observations are the "truth". As such we use the ratio between the fitted continuum in Visits 1+2 and Visit 3 to scale the Visit 3 spectrum to match the NIRCam photometry and Visit 3. 

\section*{Appendix B: ForcePho fitting}

\begin{figure*}
        \centering
	\includegraphics[width=0.85\textwidth]{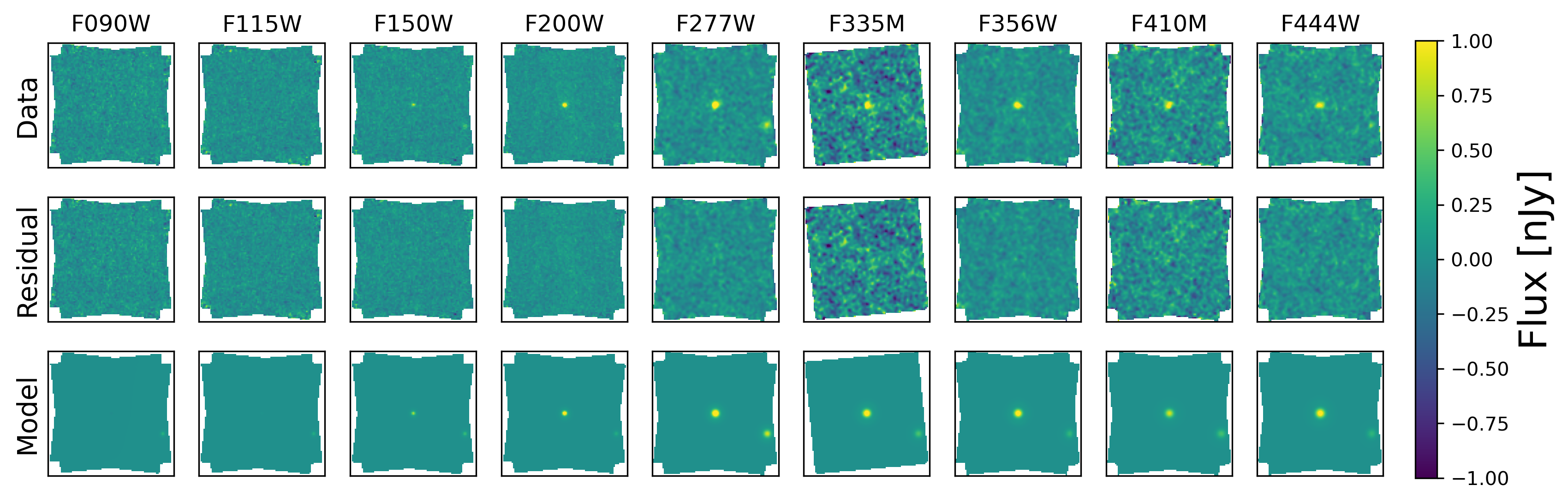}
    \caption{Forcepho morphological fitting of our sources. Top row: The 3$\times$3 arcseconds stamps for each available filter. Middle row: Residuals in each filter from our fitting. Bottom row: Best fit model for each of the filters.}
    \label{fig:forcepho}
\end{figure*}

In Figure \ref{fig:forcepho} we show the results of the Forcepho fitting across each available NIRCam filter, showing the data, residuals and the best-fit model. 


\bsp	
\label{lastpage}
\end{document}